\documentclass[prl,twocolumn,showpacs,amsmath,amssymb,floatfix, superscriptaddress]{revtex4-1}
\usepackage{graphicx}
\usepackage{dcolumn}   
\usepackage{bm}        
\usepackage{mathtools}
\usepackage{braket}
\usepackage[dvipsnames]{xcolor}
\usepackage{ulem}
\graphicspath{{./figures/}}

\usepackage[colorlinks = true,
            linkcolor = blue,
            urlcolor  = blue,
            citecolor = blue,
            anchorcolor = blue]{hyperref}
\usepackage{natbib}
\usepackage[mathscr]{euscript}
\usepackage[T1]{fontenc} 
\usepackage{listings}
\usepackage{xcolor}

\definecolor{codegreen}{rgb}{0,0.6,0}
\definecolor{codegray}{rgb}{0.5,0.5,0.5}
\definecolor{codepurple}{rgb}{0.58,0,0.82}
\definecolor{backcolour}{rgb}{0.95,0.95,0.92}

\lstdefinestyle{mystyle}{
    backgroundcolor=\color{backcolour},   
    commentstyle=\color{codegreen},
    keywordstyle=\color{magenta},
    numberstyle=\tiny\color{codegray},
    stringstyle=\color{codepurple},
    basicstyle=\ttfamily\footnotesize,
    breakatwhitespace=false,         
    breaklines=true,                 
    captionpos=b,                    
    keepspaces=true,                 
    numbers=left,                    
    numbersep=5pt,                  
    showspaces=false,                
    showstringspaces=false,
    showtabs=false,                  
    tabsize=2
}

\begin{document}
\newcommand{\rev}[1]{\textcolor{red}{#1}}
\renewcommand{\rev}[1]{#1}
	
\title{Feasible route to high-temperature ambient-pressure hydride superconductivity}

\author{Kapildeb Dolui}
\affiliation{Department of Materials Science and Metallurgy, University of Cambridge, 27 Charles Babbage Road, Cambridge CB30FS, UK}
\author{Lewis J. Conway}
\affiliation{Department of Materials Science and Metallurgy, University of Cambridge, 27 Charles Babbage Road, Cambridge CB30FS, UK}
\affiliation{Advanced Institute for Materials Research, Tohoku University, Sendai, 980-8577, Japan}
\author{Christoph Heil}
\affiliation{Institute of Theoretical and Computational Physics, Graz University of Technology, NAWI Graz, 8010 Graz, Austria}
\author{Timothy A. Strobel}
\affiliation{Earth and Planets Laboratory, Carnegie Institution for Science, 5241 Broad Branch Road, NW, Washington, DC 20015, USA}
\author{Rohit Prasankumar}
\affiliation{Intellectual Ventures, Bellevue, Washington, United States}
\author{Chris J. Pickard}
\email{cjp20@cam.ac.uk}
\affiliation{Department of Materials Science and Metallurgy, University of Cambridge, 27 Charles Babbage Road, Cambridge CB30FS, UK}
\affiliation{Advanced Institute for Materials Research, Tohoku University, Sendai, 980-8577, Japan}
\begin{abstract}

A key challenge in materials discovery is to find high-temperature superconductors.
Hydrogen and hydride materials have long been considered promising materials displaying conventional phonon-mediated superconductivity. 
However, the high pressures required to stabilize these materials have restricted their application. 
Here, we present results from high-throughput computation, considering a wide range of high-symmetry ternary hydrides from across the periodic table at ambient pressure. 
This large composition space is then reduced by considering thermodynamic, dynamic, and magnetic stability, before direct estimations of the superconducting critical temperature. 
This approach has revealed a metastable ambient-pressure hydride superconductor, Mg$_2$IrH$_6$, with a predicted critical temperature of 160\,K, comparable to the highest temperature superconducting cuprates. 
We propose a synthesis route \textit{via} a structurally related insulator, Mg$_2$IrH$_7$, which is thermodynamically stable above 15\,GPa and discuss the potential challenges in doing so. 
\end{abstract}
\maketitle		

Since Kamerlingh Onnes' discovery in 1911 of superconductivity in Mercury cooled below~4\,K\cite{Onnes1911}, a long-standing challenge in condensed matter physics has been to discover high-temperature superconductors. 
Over a century later, most practical superconductors still have critical temperatures ($T_{\rm c}$s) well below the temperature of liquid nitrogen (77\,K). 
Computational efforts to predict high-$T_{\rm c}$ materials have guided the field towards hydrides~\cite{Zurek2017, Pickard2020,Boeri2022}, typically at high pressures, where phonon-mediated BCS superconductivity is expected to play a significant role. 
Such approaches include first-principles structure prediction~\cite{Pickard2006,Pickard2007a,Duan2014,Zurek2017,Shipley2021,Chen2021}, high-throughput screening~\cite{Saha2023}, and machine-learned property prediction~\cite{Hutcheon2020,Cerqueira2023,Xie2022a,Tran2023}. 
In experiments, high-pressure hydrides of sulfur~\cite{Drozdov2015}, lanthanum~\cite{Drozdov2019}, yttrium~\cite{Troyan2021,Kong2021}, cerium~\cite{Chen2021a}, and calcium~\cite{Ma2022},
have since been synthesized and shown to have critical temperatures between 161 and 224\,K at pressures well above 100\,GPa. 

It remains a considerable challenge to predict high-$T_{\rm c}$ materials --- hydrides or otherwise --- at ambient pressure. 
Our approach leverages random structure search to generate structures, high-throughput \textit{ab-initio} property calculations to filter them, and machine-learning interatomic potentials to assess them. 
This approach becomes tractable by assuming the material will contain hydrogen, exhibit conventional phonon-mediated superconductivity, and --- as with the known superconducting high-pressure hydrides --- have a high-symmetry crystal structure containing less than around 20 atoms per unit cell. 
These assumptions allow for relatively fast searches and $T_{\rm c}$ calculations.

The aim of such a search is to identify crystal structures possessing a high $T_{\rm c}$ with thermodynamic, dynamic, and kinetic stability.
\textit{Thermodynamic} stability indicates a resilience to transformation into other phases or to decomposition into constituent species, implying the structure exists at a global energy minimum. 

This is calculated by Maxwell construction; how far a structure is from the convex hull formed of free energy and composition coordinates.~\cite{Pickard2011}
A structure on the convex hull is considered thermodynamically stable. 
A structure close to the convex hull could be described as metastable if it also possesses dynamic and kinetic stability.  
\textit{Dynamic} stability indicates a resilience to small fluctuations in atomic positions and is determined through phonon dispersion calculations. 
A structure with only positive phonon modes is considered dynamically stable. 
\textit{Kinetic} stability indicates a resilience against structural changes such as lattice distortion at higher temperatures. 
The presence of kinetic and dynamic stability implies the structure exists in a deep energy well and would be expected to be stable over a long timescale.

Since there are significantly more metastable than thermodynamically stable structures, a multitude of energetically plausible metastable high-$T_{\rm c}$ hydrides are to be expected, but comparatively few of these are actually synthesisable. 
However, many functional materials are in fact metastable states obtained through a well-designed synthesis pathway. Indeed, a recent survey of the Inorganic Crystal Structure Database found that half of all registered structures are metastable~\cite{Sun2016a}.
Given the large number of predicted metastable high-$T_{\rm c}$ hydrides, it is difficult to decide which of these are worth pursuing experimentally. This situation could improve if increased emphasis was placed on the feasibility of synthetic pathways when proposing new metastable structures.

Although most of the synthesized superconducting hydrides are only stable at impractically high pressures, synthesis routes exploiting high pressures may be viable to obtain metastable, ambient-pressure phases.
A structure may form at high pressure where it is thermodynamically stable and remain in a metastable state on recovery to lower pressures due to its dynamic and kinetic stability.
Several ternary hydrides have recently been predicted to have $T_{\rm c}$s around 70\,K, to be thermodynamically stable above 130\,GPa, and to remain dynamically stable upon decompression as low as 5\,GPa~\cite{DiCataldo2021,Lucrezi2022,Belli2022}.

\begin{figure}[t!]
\centering
\includegraphics[width=0.49\textwidth]{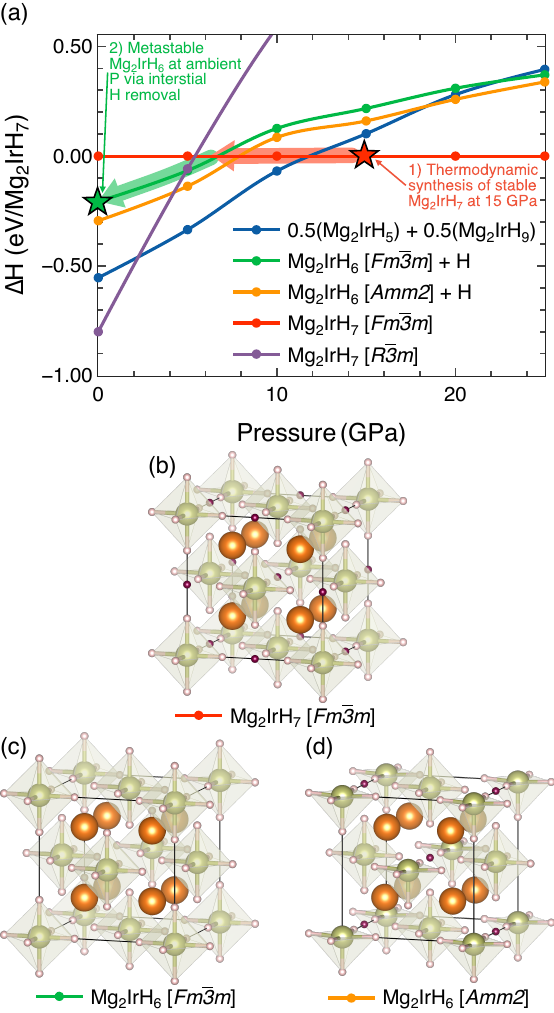}
\caption{(a) Pressure dependence of the formation enthalpies of Mg--Ir--H phases relative to the cubic Mg$_2$IrH$_7$ phase. Pressure synthesis route is indicated by arrows, starting from Mg$_2$IrH$_7$ at 15\,GPa (1) and tracking decompression to 6\,GPa, below which cubic Mg$_2$IrH$_6$ + H should form (2). (b-d) Crystal structures of Mg$_2$IrH$_7$ and Mg$_2$IrH$_6$: orange, green, pink, and maroon spheres denote the Mg, Ir, hydrido H, and interstitial H atoms, respectively.}
\label{fig:enthalpy}
\end{figure}

Here we propose an ambient-pressure high-$T_{\rm c}$ hydride; a cubic phase of Mg$_2$IrH$_6$ --- a metastable ternary hydride with $T_{\rm c}=160$\,K at ambient pressure. 
The structure consists of octahedral Ir and Mg ions occupying the 8$c$ (1/4, 1/4, 1/4) and 4$a$ (0, 0, 0) Wyckoff sites of an $Fm\Bar{3}m$ lattice with $a=6.66$\,\AA~(see Figure~\ref{fig:enthalpy}(c)). The hydrogen atoms occupy the 24$e$ (0.26, 0, 0) sites forming [IrH$_6$]$^{3-}$ hydrido clusters. This phase is less than 20\,meV/atom from the convex hull at 15\,GPa.

We propose that this phase may be recovered \textit{via} the synthesis of a second, structurally related, insulating compound, Mg$_2$IrH$_7$, which contains additional interstitial non-bonded hydrogen atoms on the 4$a$ (0.5, 0.5 0) site (maroon spheres in Figure~\ref{fig:enthalpy}(b)). This structure is thermodynamically stable above 15\,GPa. At 0\,GPa, both compositions are metastable, but the superconducting Mg$_2$IrH$_6$ configuration may be obtained by the removal of interstitial H atoms from Mg$_2$IrH$_7$. The viability of such a mechanism is reinforced by the high diffusivity of hydrogen observed in other hydrides such as LaH$_{10}$~\cite{Wang2023,Causse2023} and by our molecular dynamics calculations. 
Such a synthesis route is evocative of the concept of `remnant metastability'~\cite{Sun2016a} wherein feasible metastable materials are remnants of thermodynamically stable phases under different thermodynamic potentials. In this case, we expect Mg$_2$IrH$_6$ to be stabilized from a thermodynamically stable phase at a different pressure and chemical potential (composition).
In this Letter, we propose and discuss the synthesis route $\frac{7}{2}$\,H$_2$~+~2\,Mg~+~Ir~ $\xrightarrow[]{+15 {\rm GPa}}$  Mg$_2$IrH$_7$ $\xrightarrow[]{-15 {\rm GPa}}$ Mg$_2$IrH$_6$ $+ \frac{1}{2}$\,H$_2$.

We performed an {\it ab initio} random structure search (AIRSS)~\cite{Pickard2006, Pickard2011} at 1\,GPa for $X_aY_b$H$_c$ compositions where $a=0-3$, $b=0-3$ and $c=1-18$ with $X$ and $Y$ randomly selected from a curated element palette containing all elements up to Po, ignoring the lanthanides and noble ~\footnote{
The palette contained; Li, Be, B, C, N, O, F, Na, Mg, Al, Si, P, S, Cl, Ar, K, Ca, Sc, Ti, V, Cr, Mn, Fe, Co, Ni, Cu, Zn, Ga, Ge, As, Se, Br, Rb, Sr, Y, Zr, Nb, Mo, Tc, Ru, Rh, Pd, Ag, Cd, In, Sn, Sb, Te, I, Cs, Ba, La, Ce, Lu, Hf, Ta, W, Re, Os, Ir, Pt, Au, Hg, Tl, Pb, Bi, and Po.}.
Structures were generated with 24 or 48 symmetry operations (restricting the search to 35 cubic and hexagonal space groups~\footnote{The following space groups have 24 or 48 symmetry operations: $F\overline{4}3c$, $F\overline{4}3m$, $F4_132$, $F432$, $Fd\overline{3}$, $Fm\overline{3}$, $I\overline{4}3d$, $I\overline{4}3m$, $I4_132$, $I432$, $Ia\overline{3}$, $Im\overline{3}$, $P\overline{4}3m$, $P\overline{4}3n$, $P4_132$, $P4_232$, $P432$, $P4_332$, $P6/mcc$, $P6/mmm$, $P6_3/mcm$, $P6_3/mmc$, $Pa\overline{3}$, $Pm\overline{3}$, $Pn\overline{3}$, $Fd\overline{3}c$, $Fd\overline{3}m$, $Fm\overline{3}c$, $Fm\overline{3}m$, $Ia\overline{3}d$, $Im\overline{3}m$, $Pm\overline{3}m$, $Pm\overline{3}n$, $Pn\overline{3}m$, and $Pn\overline{3}n$ }). Geometry optimizations were performed using \textsc{castep}~\cite{Clark2005} with Perdew-Burke-Ernzerhof exchange--correlation functional~\cite{Perdew1996}, \textsc{castep} QC5 pseudopotentials, a 340\,eV plane-wave cutoff, and a $k$-point spacing of $2\pi\times~0.07$\,\AA${}^{-1}$.

The search resulted in 226,348 structures. To filter these structures for synthesizable superconductors, we perform a multi-stage, high-throughput screening process to test for thermodynamic, dynamic, and magnetic instabilities.   
Firstly, for thermodynamic stability, we calculated ternary convex hulls for all permutations of $X$+$Y$+H at 1 and at 10\,GPa (by linear extrapolation of the $PV$ term --- see section 6.4 in \cite{Pickard2011}) and retained structures within 50\,meV of the convex hull. This reduced the dataset to 1,586 structures. 
Second, for dynamic stability, we performed two iterations of \textsc{castep} phonon calculations, first on a $1\times1\times1$ $q$-grid and then $2\times2\times2$ $q$-grid, removing structures with imaginary phonon modes, reducing the dataset to 772 and then to 233 structures. 
Third, we calculated the electronic densities of states and eliminated all non-metallic structures. 
Finally, we performed a spin-polarized DFT calculation and removed any structures with non-zero spin. 
This procedure resulted in 122 structures for which we estimated $T_{\rm c}$ by calculating the electron--phonon interactions within density functional perturbation theory (DFPT) using \textsc{Quantum Espresso (QE)}~\cite{Giannozzi2020}, with a set of coarse parameters detailed in the supplementary material (SM)~\cite{suppl}, and solving the isotropic Migdal-Eliashberg (ME) equations for the Eliashberg functions, $\alpha^2 F(\omega)$. 

The results of our coarse $T_{\rm c}$ calculations is summarized in table~SIV in the SM. Of the remaining 122 structures, a common high-$T_{\rm c}$ structure type was face-centered cubic A$_2$BH$_6$. For completeness, we repeated the screening process for all 4,356 combinations of A$_2$BH$_6$ in this structure type with A and B sites occupied by atoms in our element palette.~\cite{Note1} Of these, the highest $T_{\rm c}$ calculations are shown in the SM. The face-centered cubic Mg$_2$IrH$_6$ structure had a notably high estimated $T_{\rm c}$ and a notably low formation energy and is the focus of this Letter.

The structure is shown in Fig.~\ref{fig:enthalpy}(c) and can be associated with potassium chloroplatinate, K$_2$PtCl$_6$~\cite{Ewing1928}.
Mg$_2$IrH$_6$ is also isostructural with a family of insulating ternary hydrides, including Mg$_2$(Fe,Ru,Ni,Os)H$_6$ and K$_2$SiH$_6$, which have emerged as appealing energy storage or electrode materials~\cite{Huang1991,Raman2002,Zaidi2013,Xie2022}.

To investigate the overall stability of Mg$_2$IrH$_6$, we carried out extensive searches on the Mg--Ir--H ternary system using an Ephemeral Data Derived Potential (EDDP)~\cite{Pickard2022}, trained by a similar method as used for the Lu--N--H system~\cite{Ferreira2023}. 
The most stable structures predicted by the EDDP, along with all Mg--Ir--H structures obtained from the Materials Project~\cite{Jain2013}, were re-optimized at a range of pressures up to 40\,GPa by \textsc{castep} using PBE and PBEsol exchange--correlation functionals. 
For these calculations, we used a 600\,eV plane-wave cutoff, a $k$-point spacing of $2\pi\times~0.03$\,\AA${}^{-1}$, and \textsc{castep} C19 pseudopotentials.

\begin{figure}[t!]
\centering
\includegraphics[width=0.49\textwidth]{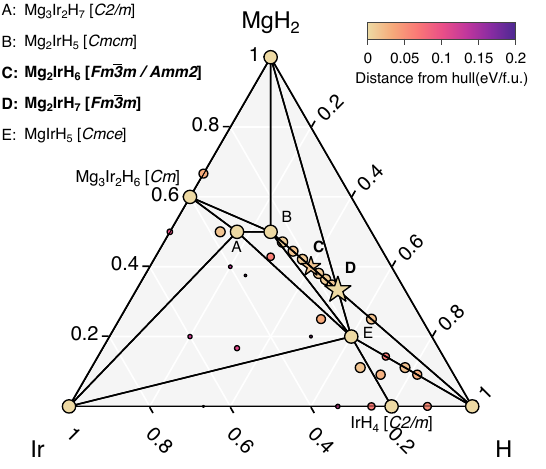}
\caption{Ternary convex hull at 20\,GPa, calculated with PBEsol. Black lines are ridges connecting thermodynamically stable points. Star symbols indicate Mg$_2$IrH$_7$ and Mg$_2$IrH$_6$. Only structures within 200\,meV of the convex hull are shown. Symbols are smaller (and more blue) with increasing distance from the convex hull. At 20\,GPa, before considering quantum nuclear effects, the $Amm2$ phase of Mg$_2$IrH$_6$ is more stable than $Fm\Bar{3}m$.}
\label{fig:ternary}
\end{figure}
Figure~\ref{fig:ternary} shows the resulting convex hull at 20\,GPa, where Mg$_2$IrH$_6$ is present but around 10\,meV/atom above the convex hull. Also present are insulating phases of Mg$_2$IrH$_5$, Mg$_2$IrH$_9$, as well as cubic  and rhombohedral phases of Mg$_2$IrH$_7$.
The rhombohedral phase of Mg$_2$IrH$_7$ is on the convex hull at 0~GPa. Above 15\,GPa, the cubic phase of Mg$_2$IrH$_7$ is on the hull.

Figure~\ref{fig:enthalpy}(a) shows the pressure evolution of cubic and rhombohedral Mg$_2$IrH$_7$ and competing decomposition routes. The thick lines and arrows indicate a proposed synthesis route starting from cubic Mg$_2$IrH$_7$ at 15\,GPa, followed by decompression to 6\,GPa, below which Mg$_2$IrH$_6$ + H is favorable down to 0\,GPa.

We also performed a comprehensive analysis of alternative decomposition routes from cubic Mg$_2$IrH$_7$ at 0\,GPa by enumerating all combinations of Mg$_2$IrH$_x$ for $5\le x\le7$ formed by removing hydrogen atoms starting from the conventional unit cell shown in Figure~\ref{fig:enthalpy}(b). This enumeration process yielded 35,178 symmetrically inequivalent structures. The single-point energy of each structure was calculated using \textsc{castep}, after which full geometry optimizations were performed for the lowest-energy structures. Most permutations of hydrogen vacancies result in a lowering of symmetry and small lattice distortions. The pseudo-binary convex hull of these permutations is shown in Figure~S4 in the SM. Although the $Fm\Bar{3}m$ phase of Mg$_2$IrH$_6$ is one of the lower-energy configurations, we note the presence of even lower-energy configurations in which some of the IrH$_6$ clusters are broken into IrH$_5$ clusters and interstitial H atoms. The lowest-energy structure of this type is an insulating orthorhombic ($Amm2$) phase, shown in Figure~\ref{fig:enthalpy}(d) and S6 in the SM, with a $c/a$ ratio of 1.03.

This presents a challenge in the synthesis route; to obtain a superconducting rather than insulating phase, the hydrogen content must be controlled while preserving the IrH$_6$ clusters. 

To understand the nature of the kinetic stability of these phases, we performed molecular dynamics calculations using an EDDP on a range of Mg$_2$IrH$_x$ compounds at 0GPa and up to 1200\,K (see SM, Figures~S2-3).  Notably, cubic Mg$_2$IrH$_7$ has the lowest activation energy (0.06\,eV) for hydrogen diffusion. Of the hydrogen-poorer structures, those with more fully occupied IrH$_6$ clusters, such as cubic Mg$_2$IrH$_6$ (0.56\,eV), had a higher activation energy than those with partially occupied clusters such as orthorhombic Mg$_2$IrH$_6$ (0.26\,eV).

Moreover, above 600\,K the total number of fully occupied clusters tends to increase throughout the trajectory. This agrees well with our \textit{ab-initio} quasi-harmonic free energy calculations (see Figure~S9 in the SM). At 300\,K there is very little diffusion.

We can infer from these results that cubic Mg$_2$IrH$_6$ possesses a relatively high degree of kinetic stability and that it may be feasible for the interstitial hydrogen atoms to diffuse out of the system. Since at 300\,K and lower, there is very little diffusion we can expect the structure to persist after cooling.

This indicates a complex synthesis route involving heating, cooling, compression and decompression may be required to obtain the superconducting phase from the more hydrogen-rich insulating phase.

Of the many experimental observables associated with a metal--insulator transition, we specifically note that, compared to the cubic structure, orthorhombic Mg$_2$IrH$_6$ exhibits a hardened Raman-active Ir--H stretching mode as shown in Figure~S10 in the SM.
Compared with cubic Mg$_2$IrH$_7$, cubic Mg$_2$IrH$_6$ exhibits a softened $T_{2g}$ mode associated with Ir--H bending.

\begin{figure}[t!]
\centering
\includegraphics[width=0.49\textwidth]{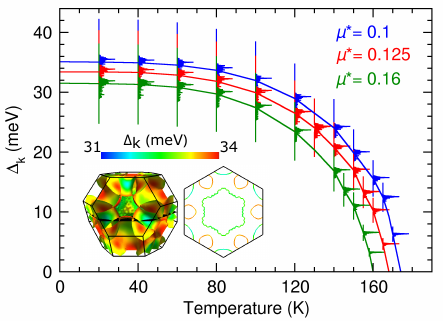}
\caption{Histograms of the energy-dependent distribution of the anisotropic superconducting gap $\Delta_{\mathbf{k}}$ on the Fermi surface of Mg$_2$IrH$_6$, evaluated for each temperature solving the ME equations with $\mu^*=$ 0.1 (blue), 0.125 (red), and 0.16 (green). The solid dashed lines are guides to the eye tracking the average $\Delta_{\mathbf{k}}(T)$. The inset shows the 3D Fermi surface (left) and 00$\overline{1}$ projection (right) colored according to the $k$-dependent gap values $\Delta_{\mathbf{k}}$ obtained from solving the anisotropic ME equation at 15\,K with $\mu^*$= 0.125.}
\label{fig:gap}
\end{figure}

Having identified a promising and potentially accessible metastable high-$T_{\rm c}$ structure, we performed robust electron--phonon coupling calculations on denser, well-converged, Wannier interpolated, $k$- and $q$-meshes using \textsc{QE}~\cite{Giannozzi2020} and \textsc{EPW}~\cite{Giustino2007,Margine2013,Ponce2016}.
The superconducting gap was then calculated by solving the anisotropic ME equations. The details of these calculations are provided in the SM. 

\begin{figure}[t!]
\centering
\includegraphics[width=0.48\textwidth]{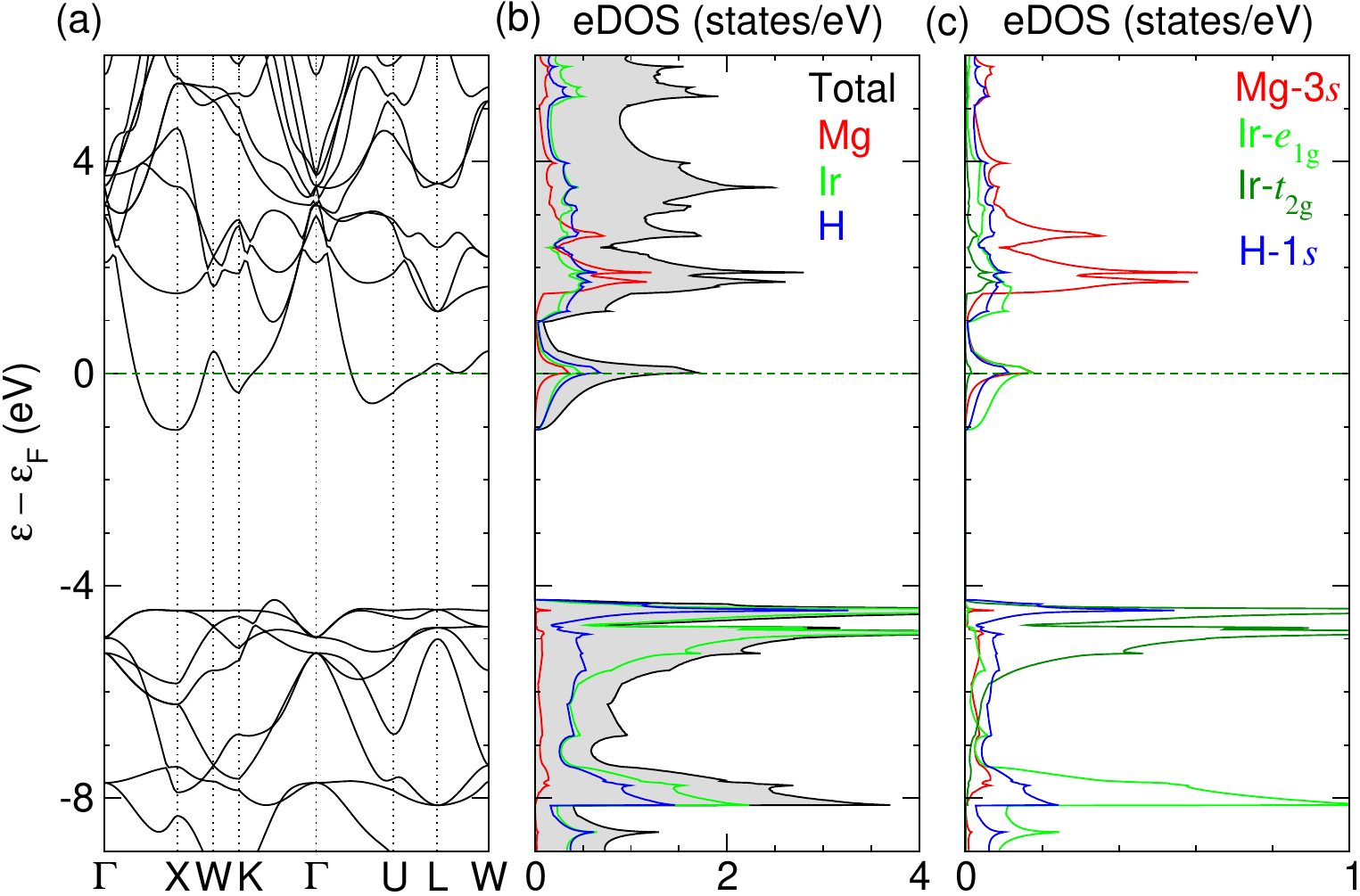}
\caption{(a) Electronic band structure of $Fm\overline{3}m$ Mg$_2$IrH$_6$ at
ambient pressure. (b) The total electron density of states (eDOS) projected onto Mg (red), Ir (green), and H (blue) atoms. (c) The eDOS is projected onto Mg-3$s$ (red), Ir-$e_{g}$ (light green), Ir-$t_{2g}$ (dark green) and H-1$s$ orbitals (blue). Dark green dashed line indicates the Fermi level.}
\label{fig:eband}
\end{figure}

Figure~\ref{fig:gap} shows the superconducting gap distribution as a function of temperature for a range of $\mu^*$ values.
The calculated $T_{\rm c}$ is between 160\,K and 175\,K at ambient pressure.
This high $T_{\rm c}$ can be rationalized by examining the density of states at the Fermi level in Figure~\ref{fig:eband}, where there is a van-Hove-like singularity (VHS) caused by flat bands along the L--W high-symmetry path.
Correlation effects are unlikely to significantly change the calculated $T_{\rm c}$. 
The VHS and the phonon dispersion are not significantly modified by the inclusion of a Hubbard parameter, $U=4$\,eV, on the Ir \textit{d}-orbitals (See SM Figure~S8).

\begin{figure}[t]
\centering
\includegraphics[width=0.48\textwidth]{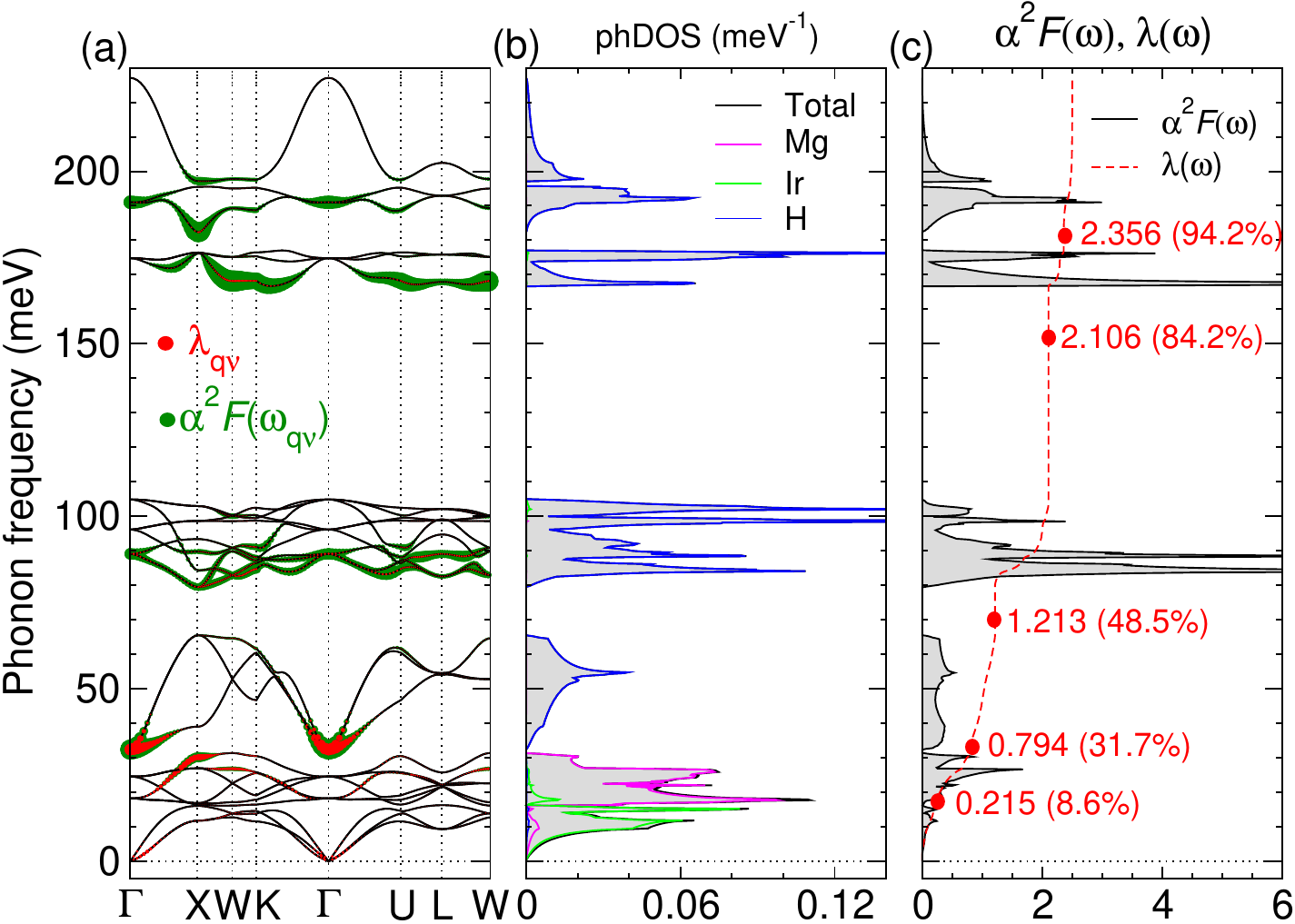}
\caption{(a) Phonon dispersion along a high-symmetry path in the Brillouin zone. The radius of the red and green circles in the panel (a) is proportional to the magnitude of $\lambda_{{\rm \mathbf{q}}\nu}$ and $\alpha^2F(\omega_{{\rm \mathbf{q}}\nu}$) for each phonon mode, respectively. (b) Total (black) and projected phonon densities of states (phDOS) onto Mg (magenta), Ir (green), and H (blue) atoms. (c) Isotropic Eliashberg spectral function, $\alpha^2F(\omega)$, and cumulative frequency-dependent electron-phonon coupling strength $\lambda(\omega)$. The labelled red circles indicate the running total (and fraction) of $\lambda(\omega)$.}
\label{fig:phband}
\end{figure}

Figure~\ref{fig:phband}(c) shows the isotropic Eliashberg spectral function, $\alpha^2F\rm (\omega\rm)$, and the cumulative electron--phonon coupling parameter, $\lambda(\omega)$, obtained by integrating $\alpha^2F\rm (\omega\rm)/\omega$ over $\omega$. This gives a total $\lambda$ of about 2.5 for Gaussian smearing widths of 0.005 Ry. In our fully converged anisotropic \textsc{EPW} calculations, $\lambda$ reduces to 2.3.

The projected densities of states in Figure~\ref{fig:eband}(b) and (c) show a significant contribution from H-1$s$ orbitals, enabling the high-frequency phonon modes to couple with electrons near the Fermi surface.
Mode-resolved contributions to $\lambda$, as shown in Figure~S7 of the SM, indicate that the molecular modes of the IrH$_6$ clusters are the most significant. As such, we see a significant contribution to $\lambda$ from hydrogen phonon bands at around 100 and 200\,meV. 

Quantum anharmonic effects can significantly impact vibrational properties, particularly in hydrides. 
To assess their importance for cubic Mg$_2$IrH$_6$, we employed the stochastic self-consistent harmonic approximation (SSCHA)~\cite{Monacelli2021}. 
We observe that the vibrational modes of Ir and Mg are almost unchanged when going from the harmonic approximation to the anharmonically corrected dispersion (see Figure~S12 in the SM). 
The low-frequency H-$T_{1g}$ modes between 30--80 meV harden, while the high-frequency H-$T_{1u}$, -$E_{g}$, -$A_{1g}$ modes above 150 meV soften, almost equally in magnitude by 15 meV. 
We attribute the hardening of H modes to the effects of anharmonicity, while the softening is mainly due to the inclusion of quantum ionic motion, as has been observed in other low-pressure hydride materials~\cite{Lucrezi2022a}. 
These calculations show that the inclusion of quantum anharmonic effects does not significantly alter the phonon dispersion of cubic Mg$_2$IrH$_6$. 

To understand the chemical bonding of cubic Mg$_2$IrH$_6$, we calculated the electron localization function (ELF)~\cite{Becke1990} which is plotted in the SM (Figure~S5). The ELF shows isolated proto-spherical isosurfaces surrounding the hydrogen atoms illustrating the partially ionic character of the Ir--H bonds, originating from the large difference in electronegativity between Ir and H.

A large gap below the Fermi level in the electronic density of states (Figure~\ref{fig:eband}b) indicates that hole doping would result in the formation of a wide-gap insulator. 
For Mg$_2$IrH$_6$, hybridization between H-1$s$ orbitals and Ir-$e_{1g}$ orbitals results in splitting the bonding--antibonding states, which contributes to a significant crystal field splitting of $\sim$4\,eV between weakly hybridized occupied Ir-$t_{2g}$ states and unoccupied Ir-$e_{1g}$ states. Therefore, if the Ir site is instead occupied by an atom with one fewer $d$ electron (i.e., $d^6$ valency elements such as Fe, Ru and Os) the material would exhibit a band gap within the $d$ manifold, consistent with well-known insulators and our screening of this structure type for high-$T_{\rm c}$ compounds. Therefore, total or partial substitution of Ir with other transition metals, or adjustment of the overall H stoichiometry, may significantly impact the bonding nature and the electronic structure, leading to a variety of possible outcomes including the formation of insulators, ferromagnets, or perhaps even higher-$T_{\rm c}$ superconductors.

In this Letter, we have presented a promising result from a wide search for hydrides exhibiting phonon-mediated BCS superconductivity. The tendency for electrons to form Cooper pairs --- the mechanism behind BCS superconductivity --- must compete with other instabilities resulting from electron--electron and electron--phonon coupling; magnetism, charge-density waves, disorder, and structural distortions.
By employing \textsc{AIRSS} across the periodic table, high-throughput calculations, and DFPT, we have filtered a wide compositional and structural space (albeit a high-symmetry subspace) to extract candidates with only the Cooper pair instability. 
In doing so, we have predicted the existence of cubic Mg$_2$IrH$_6$, an ambient-pressure, phonon-mediated, superconducting hydride with a $T_{\rm c}$ of about 160\,K. 
We demonstrate using machine-learning potentials that there may exist a viable high-pressure synthesis route \textit{via} an intermediate cubic Mg$_2$IrH$_7$ phase, thermodynamically stable above 15\,GPa. 
The extraction of interstitial hydrogen from Mg$_2$IrH$_7$ represents a plausible route to create superconducting Mg$_2$IrH$_6$, although kinetic barriers between competing phases warrant careful consideration.
This work demonstrates computationally the feasibility of high-$T_{\rm c}$ hydrides at ambient-pressure and reflects a cautious but positive outlook for hydride superconductivity. 

{\it Acknowledgments.} We thank Warren Pickett and Eva Zurek for their insightful discussion. This work was supported by the Deep Science Fund at Intellectual Ventures.

{\it Note} In an independent computational search, Sanna et {\it al.}~\cite{Sanna2024} have also indentified the possibility of high temperature conventional superconductivity in Mg$_2$IrH$_6$ and related compounds, to be compared with Tables IV and V of our SM. Likely due to their less focussed structure searches, they predicted Mg$_2$IrH$_6$ to be thermodynamically stable. The wide range in predicted $T_{\rm c}$, from around 65 to 160 K, highlights the sensitivity of computing $T_{\rm c}$ using current methodologies, and for Mg$_2$IrH$_6$ especially so due to its exceptional electronic structure.

\clearpage 

\onecolumngrid
\begin{center}
\textbf{Supplemental Material for ``Feasible route to high-temperature ambient-pressure hydride superconductivity"}
\end{center}

\begin{itemize}
\item Computational Methods
\begin{itemize}
    \item `Coarse' parameters used in DFPT calculations for high-throughput search
    \item `Robust' parameters used in $T_{\rm c}$ calculations for Mg$_2$IrH$_6$
    \item Anisotropic superconducting gap calculations
    \item Anharmonic phonon calculations
    \item Quasiharmonic free energy calculations
    \item Raman Calculations
    \item Molecular Dynamics
\end{itemize}
\item Simulated X-ray diffraction patterns
\item Molecular dynamics simulation
\item Defect calculations
\item Electronic structures of $Fm\bar{3}m$ and $Amm2$ phases of Mg$_2$IrH$_6$
\item Phonon mode-resolved $\lambda$
\item Electronic and phonon dispersion calculation using DFT+$U$
\item Free energy calculations
\item Simulated Raman
\item Anharmonic phonon dispersion
\item Crystal Structures
\item Full Convex Hull
\item Elastic constants
\item Coarse $T_{\rm c}$ calculations
\begin{itemize}
    \item High symmetry search
    \item A$_2$BH$_6$ high throughput Screening
\end{itemize}

\end{itemize}
\clearpage
\section{Computational Methods}
\subsection{\bf `Coarse' parameters used in DFPT calculations for high-throughput search}
For the coarse DFPT calculations using \textsc{Quantum Espresso}, we used the PBE exchange-correlation functional in combination with scalar-relativistic ultrasoft Vanderbilt pseudopotential and energy cutoffs for the plane-waves and change density of 50 and 500 Ry, respectively. We used a $q$-mesh of $2\times 2\times 2$ and $k$-mesh of $4\times 4\times 4$ with a phonon self-convergence threshold of 10$^{-16}$. Electron--phonon coupling elements were calculated using a dense $k$-mesh of $8\times 8\times 8$. The numeric implementation of the electron--phonon coupling elements requires Dirac delta functions to be replaced with Gaussians with a given width. \rev{We chose to calculate $T_{\rm c}$ for a range of gaussian broadening and rank by both the smallest and largest broadening values to account for the poor convergence.}

\subsection{\bf `Robust' parameters used in $T_{\rm c}$ calculations for Mg$_2$IrH$_6$}
For the DFPT~\cite{Baroni2001} calculations, we used the PBE exchange-correlation functional~\cite{Perdew1996} in combination with scalar-relativistic ultrasoft Vanderbilt pseudopotentials~\cite{Garrity2014} and energy cutoffs for the plane-waves and change density of 80 and 640 Ry, respectively. We used a $q$-mesh of $6\times6\times 6$ and $k$-mesh of $18\times18\times18$ with a phonon self-convergence threshold of 10$^{-16}$ and a Methfessel-Paxton type smearing~\cite{Methfessel1989} of 0.04 Ry for the Brillouin zone integration. 

The \textsc{EPW}~\cite{Ponce2016} code was used to interpolate electronic and vibrational properties employing maximally localized Wannier functions. For this, we performed a further static DFT calculation on a $6\times~6\times6$ k-grid and Wannierize the bands using the selected columns of the density matrix (SCDM)~\cite{Damle2015} method using {\it error} function with $\mu$= -4.948\,eV and $\sigma$ = 6.804\,eV, applied in the energy range of -80 to 16\,eV around the Fermi level. We interpolated the bands and phonons both onto $36\times36\times36$ grid.

\subsection{\bf Anisotropic superconducting gap calculations}
The superconducting gap function was calculated solving the anisotropic Eliashberg theory, where the key equations~\cite{Ponce2016, Margine2013} are:
\begin{eqnarray}
Z({\bf k},i\omega_n) =
1 + \frac{\pi T}{N_{\rm F}\omega_n} \sum_{{\bf k}' n'}
\frac{ \omega_n' }{ \sqrt{\omega_n'^2+\Delta^2({\bf k}',i\omega_n')} } 
\times\lambda({\bf k},{\bf k}',n\!-\!n') \delta(\epsilon_{{\bf k}'}),
\end{eqnarray}
and 
\begin{eqnarray}
Z({\bf k},i\omega_n) \Delta({\bf k},i\omega_n) =
\frac{\pi T}{N_{\rm F}} \sum_{{\bf k}' n'} \frac{ \Delta({\bf k}',i\omega_n') }{ \sqrt{\omega_n'^2+\Delta^2({\bf k}',i\omega_n')} } 
\times\left[ \lambda({\bf k},{\bf k}',\!n-\!n')-N_{\rm F} V({\bf k}-{\bf k}')\right] \delta(\epsilon_{{\bf k}'}),
\end{eqnarray}
Here $Z, \Delta, i\omega_{n}, T, N_{{\rm F}}, V$, and $\mathbf{k}$ are the electron-phonon renormalization function, the superconducting gap, $n^{\rm th}$ fermionic Matsubara frequencies, the absolute temperature, the density of electronic states at the Fermi level, screened Coulomb interaction and the wavevector index, respectively; the Dirac delta function is smeared by the Lorenzian function; and the anisotropic electron-phonon coupling $\lambda$ is given by, 
\begin{equation}
\lambda({\bf k},{\bf k}',n - n') = \int_{0}^{\infty} d\omega
\frac{2\omega}{(\omega_n - \omega_n')^2+\omega^2}\alpha^2F({\bf k},{\bf k}',\omega).
\end{equation}
By solving the Eq. 1 and 2 numerically for each temperature separately, anisotropic ME superconducting transition temperature $T_{\rm c}^{\rm ME}$ can be obtained when the $\Delta(i\omega) \rightarrow 0$. Within the \textsc{EPW} code, the ME equations are solved including the electronic states around $\pm0.7$ eV of the Fermi level, and the Dirac $\delta$ functions are replaced by Lorentzians of widths 25 and 0.1 meV for electrons and phonons, respectively. The Matsubara frequency cutoff is set to 2.5 eV, which is approximately ten times greater than the highest phonon frequency. The static screened Coulomb interaction $V({\bf k}-{\bf k}')$ is replaced by the standard value of Morel-Anderson pseudopotential $\mu^*$, which is materials dependent and ranges typically from 0.1-0.2~\cite{Margine2013}.\\

\subsection{\bf Anharmonic phonon calculations}

The computations within the SSCHA framework are performed using the constant-volume relaxation mode within a 2$\times$2$\times$2 supercell. This involves minimizing the free energy by adjusting the average atomic positions ($\mathcal{R}$) and the force constants ($\Phi$), which is implemented through the SSCHA python package \cite{Monacelli2021}. Our initial estimates for $\mathcal{R}$ and $\Phi$ are based on the equilibrium atomic positions from DFT and the DFPT dynamical matrices, respectively. These are obtained on a 2$\times$2$\times$2 $\mathbf{q}$-grid.

The DFT calculations provide us with the total energies, forces, and stress tensors for individual cases. After each minimization iteration, we generate a new population with an increased number of individuals $N$ using the minimized trial density matrix until convergence is achieved. We set two criteria for stopping the minimization loops: firstly, a Kong-Liu ratio, which evaluates the effective sample size~\cite{Note4} and should reach a value of 0.2, and secondly, a ratio of less than $10^{-7}$ between the free energy gradient with respect to the auxiliary dynamical matrix and its stochastic error.

To obtain anharmonic phonon dispersions, we analyze positional free-energy Hessians, including the fourth-order term. The final atomic positions are determined from the converged average atomic positions $\mathcal{R}$, and the pressure is obtained as the derivative of the converged free energy with respect to a strain tensor. All calculations are conducted at a temperature of zero kelvin.

\subsection{Quasiharmonic free energy calculations}

We used \textsc{phonopy}\cite{Togo2023} to calculate the free energy as a function of pressure and temperature. Phonon calculations were carried out using the finite-displacement method on a $2\times2\times2$ supercell of the conventional unit cells of $Fm\bar{3}m$ and $Amm2$ structures. 
Forces and energies were calculated with \textsc{castep}.
For these calculations, we used a 600\,eV plane-wave cutoff, a $k$-point spacing of $2\pi\times~0.03$\,\AA${}^{-1}$, and \textsc{castep} C19 pseudopotentials.

\subsection{Raman Calculations}
We used \textsc{castep} to calculate the gamma-point vibrational frequencies and intensities (for insulating structures) using linear-response theory. For these calculations, we used a 600\,eV plane-wave cutoff, a k-point spacing of $2\pi\times~0.03$\,\AA${}^{-1}$, and \textsc{castep} C19 pseudopotentials.

\subsection{Molecular Dynamics}

\rev{For molecular dynamics, we trained a second EDDP, this time using `markers', known crystal structures which are shaken and included in the training dataset (see ref~\cite{Pickard2022}). We then iteratively added data using the usual iterative relax-shake-train procedure on compositions between Mg$_2$IrH$_{5-5}$ and Mg$_2$IrH$_{7}$. }

\rev{A molecular dynamics code, \texttt{ramble} (see ~\cite{Salzbrenner2023}), is designed to work well with EDDPs. We used \texttt{ramble} to calculate MD trajectories at 0\,GPa. Simulation cells were made from supercells chosen using a `nearly cubic' supercell technique outlined by Salzbrenner \textit{et al.}\cite{Salzbrenner2023} and contained approximately 1,000 atoms. We ran NpT calculations at 100, 300, 600, 900 and 1200\,K all at 0\,GPa for 50\,ps with a timestep of 0.05\,fs. We measured diffusion coefficients from the mean-squared displacement after $\sim$20\,ps of initialisation.}

\clearpage
\section{Simulated X-ray Diffraction Patterns}
\begin{figure}[h!]
    \centering
    \includegraphics[width=0.8\linewidth]{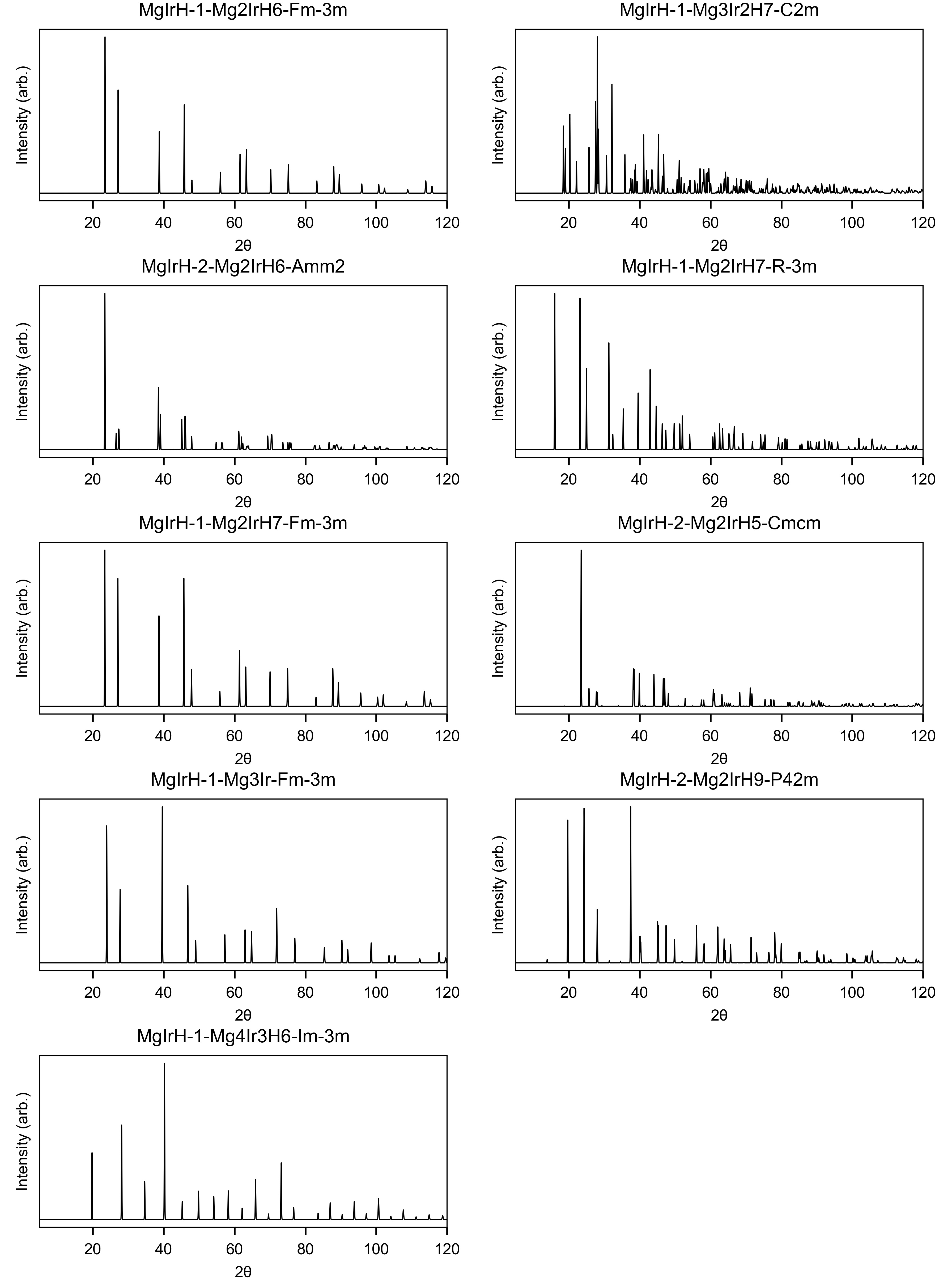}
    \caption{Simulated XRD patterns (Cu$_{k \alpha}$ radiation\rev{, 8.0\,keV}) for cubic Mg--Ir--H structures at 0\,GPa.}
    \label{fig:xrd}
\end{figure}
\clearpage

\section{Molecular Dynamics Simulations}

\begin{figure}[hb]
    \centering
    \includegraphics[width=0.8\linewidth]{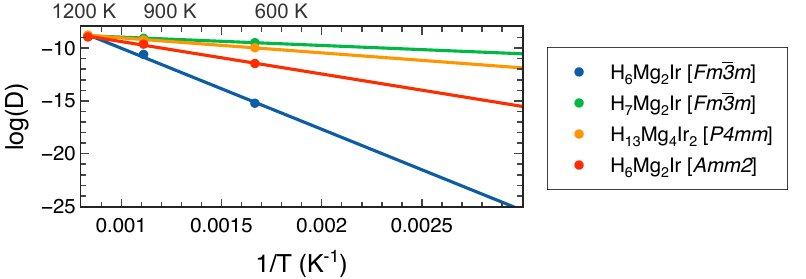} 
    \caption{Diffusion coefficient against inverse temperature. Points are calculated values; lines are linear fits.}
    \label{fig:Arrhenius}
\end{figure}

\rev{In figure~\ref{fig:Arrhenius}, we plot the log of the diffusion coefficient for the hydrogen atoms, $D$, as a function of inverse temperature. A linear fit is used to extract the activation energy, $E_a$, from the Arrhenius relationship, $D = A \exp{(\frac{-E_a}{k_BT})}$. The activation energies are shown in table~\ref{tab:Arrhenius}; The cubic phase of Mg$_2$IrH$_6$ has the largest activation energy.}

\begin{table}[b!]
    \centering
    \caption{Activation Energies}
    \begin{tabular}{ll} \hline \hline
Structure   & Activation Energy     \\ \hline
1-H6Mg2Ir-Fm-3m   & 0.56\,eV \\
1-H7Mg2Ir-Fm-3m   & 0.06\,eV \\
2-H13Mg4Ir2-P4mm  & 0.11\,eV \\
2-H6Mg2Ir-Amm2    & 0.26\,eV \\ \hline\hline
    \end{tabular}
    \label{tab:Arrhenius}
\end{table}

\begin{figure}[hb]
    \centering
    \includegraphics[width=\linewidth]{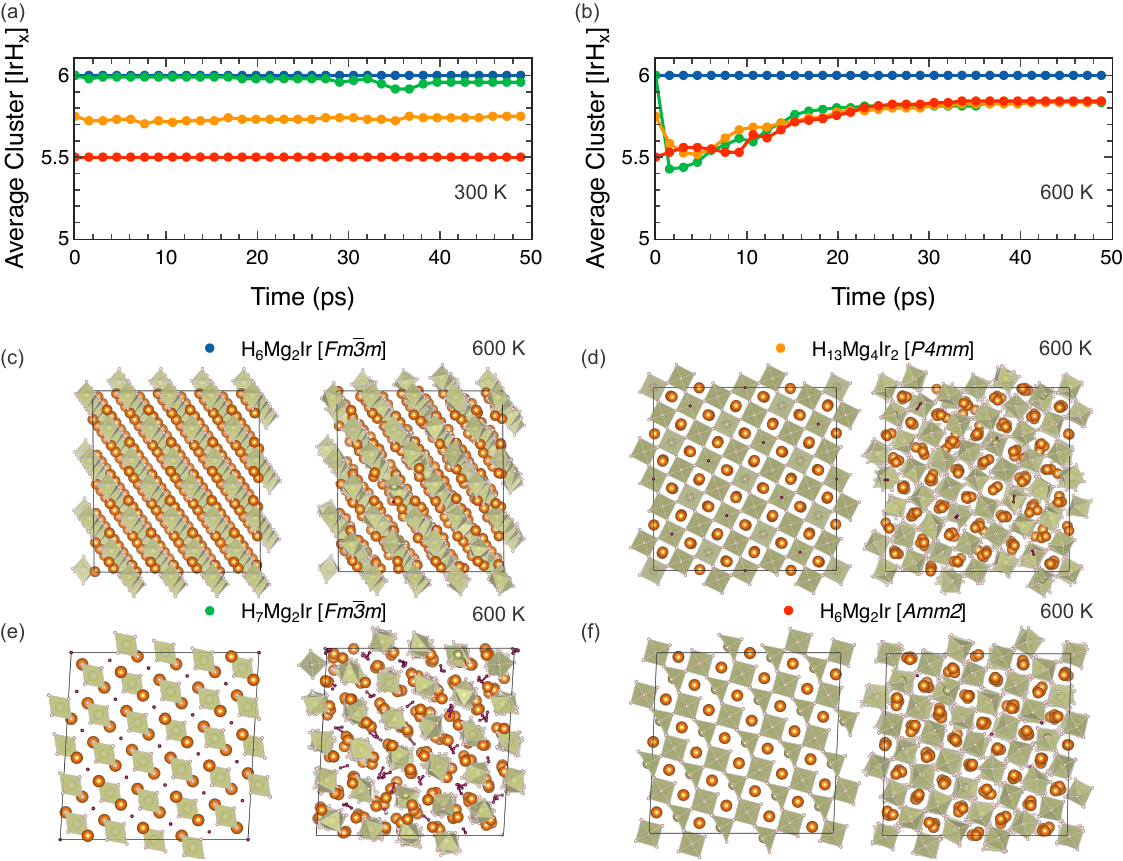} 
    \caption{(a-b) Ir-H cluster sizes during the MD trajectories. (c-f) Snapshots of the trajectories at 0 (left) and 50\,ps (right). Ir-H clusters are shown by polyhedra, Mg atoms by orange spheres. Hydrogen atoms not in an Ir-H cluster are shown by maroon spheres.}
    \label{fig:Clusters}
\end{figure}

\rev{Throughout the MD trajectory, we monitor the average size of IrH$_x$ clusters in Figure~\ref{fig:Clusters}, defined by Ir-H distances less than 1.9\,\AA~over a 0.05\,ps rolling-average. For example, the ground-state cubic phase of Mg$_2$IrH6 has an average Ir-H cluster of IrH$_6$, whereas the $Amm2$ structure has 50\% IrH$_6$ clusters and 50\% IrH$_5$ clusters, and hence an average of IrH$_{5.5}$. Figure~\ref{fig:Clusters} also shows snapshots at 0 and 50 ps of the MD trajectory. Hydrogen atoms in an Ir-H cluster are light pink, otherwise they are maroon. }

\rev{At 300\,K, Figure~\ref{fig:Clusters}(a), the clusters remain intact, with occasional site hopping. However, at 600\,K, Figure~\ref{fig:Clusters}(b), there is more frequent site hopping. In cubic Mg$_2$IrH$_6$, site hopping is between different IrH$_6$ clusters, \textit{via} the interstitial site, hence the average cluster size remains constant. In cubic Mg$_2$IrH$_7$, there is a much lower activation energy for H diffusion. The diffusion also allows for H$_2$ formation, any remaining H preferentially fill IrH$_6$ clusters. A slower process of H$_2$ formation and decomposition is also observed, meaning at very long timescales, the diffusion may persist until all clusters are filled.}

\begin{figure}[hb]
    \centering
    \includegraphics{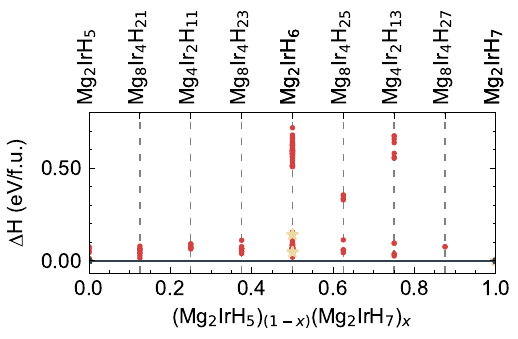}
    \caption{Convex hull of low-energy structures listed in table~\ref{tab:defects} after DFT relaxation at 0 GPa. \rev{1~f.u.=(Mg$_2$IrH$_5$)$_x$(Mg$_2$IrH$_7$)$_{1-x}$  } Stars indicate the $Amm2$ and $Fm\bar{3}m$ Mg$_2$IrH$_6$ and $Fm\bar{3}m$ Mg$_2$IrH$_7$.}
    \label{fig:defects_hull}
\end{figure}

\rev{Mg$_4$Ir$_2$H$_{13}$ is a `defected' structure from Figure~\ref{fig:defects_hull}. In the ground-state it has an average Ir cluster size of IrH$_{5.75}$ and several `non bonded' hydrogen atoms. Similarly, this composition eventually forms mostly fully occupied IrH$_6$ clusters and H$_2$. However, we also observe a gradual decomposition of H$_2$ molecules.}

\rev{Mg$_2$IrH$_6$ [$Amm2$] in the ground-state has a mixture of IrH$_5$ and IrH$_6$ clusters and `non-bonded' hydrogen atoms. It has a tendency, above 600\,K, to reduce the number of IrH$_5$ clusters and to form more IrH$_6$ clusters, as in Mg$_2$IrH$_6$ [$Fm\bar{3}m$]. }

\rev{An interpretation of these results is that the cubic phase of Mg$_2$IrH$_6$ could be synthesised by heating, to drive the system towards containing more IrH$_6$ clusters, followed by quenching to preserve the structure.}


\clearpage

\section{Defect calculations}
\begin{table}[hb]
    \centering
    \begin{ruledtabular}
    \begin{tabular}{llll}
Formula        & Vacancies &  Structures &  Symmetrically Inequivalent Structures \\
\hline 
H$_{7}$Mg$_{2}$Ir        & 0 &           1 &          1\\
H$_{27}$Mg$_{8}$Ir$_{4}$ & 1 &          28 &          2\\ 
H$_{13}$Mg$_{4}$Ir$_{2}$ & 2 &         378 &         12\\ 
H$_{25}$Mg$_{8}$Ir$_{4}$ & 3 &       3,276 &         53\\ 
H$_{6}$Mg$_{2}$Ir        & 4 &      20,475 &        252\\ 
H$_{23}$Mg$_{8}$Ir$_{4}$ & 5 &      98,280 &        927\\ 
H$_{11}$Mg$_{4}$Ir$_{2}$ & 6 &     376,740 &      3,158\\ 
H$_{21}$Mg$_{8}$Ir$_{4}$ & 7 &   1,184,040 &      8,875\\ 
H$_{5}$Mg$_2$Ir          & 8 &   3,108,105 &     21,899\\ 
    \end{tabular}
    \end{ruledtabular}
    \caption{Enumeration of symmetrically equivalent hydrogen-deficient structures starting from the conventional unit cell of Mg$_2$IrH$_7$.}
    \label{tab:defects}
\end{table}
\clearpage

\section{Electronic structures of $Fm\bar{3}m$ and $Amm2$ phases of Mg$_2$IrH$_6$}
\begin{figure}[h!]
\centering
\includegraphics[width=0.99\textwidth]{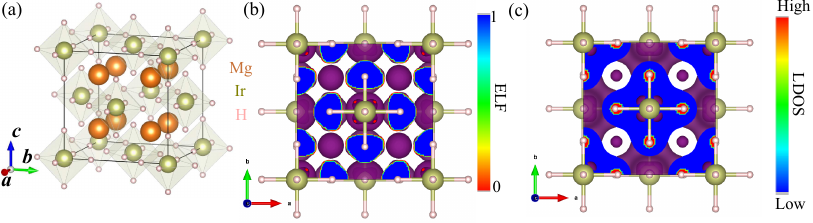}
\caption{ (a) Crystal structure of $Fm\overline{3}m$ Mg$_2$IrH$_6$ at ambient pressure. (b) Isosurface of the Electron localization function at ELF=0.4. The slice on the 001 plane at $z$=0 is also shown by colors indicated by the color bar. (c) Isosurface of the local density of states (LDOS) at the Fermi level. Slice also on the 001 plane at $z$=0.}
\label{fig:ELF}
\end{figure}
\begin{figure}[hb]
\centering
\includegraphics[width=0.75\textwidth]{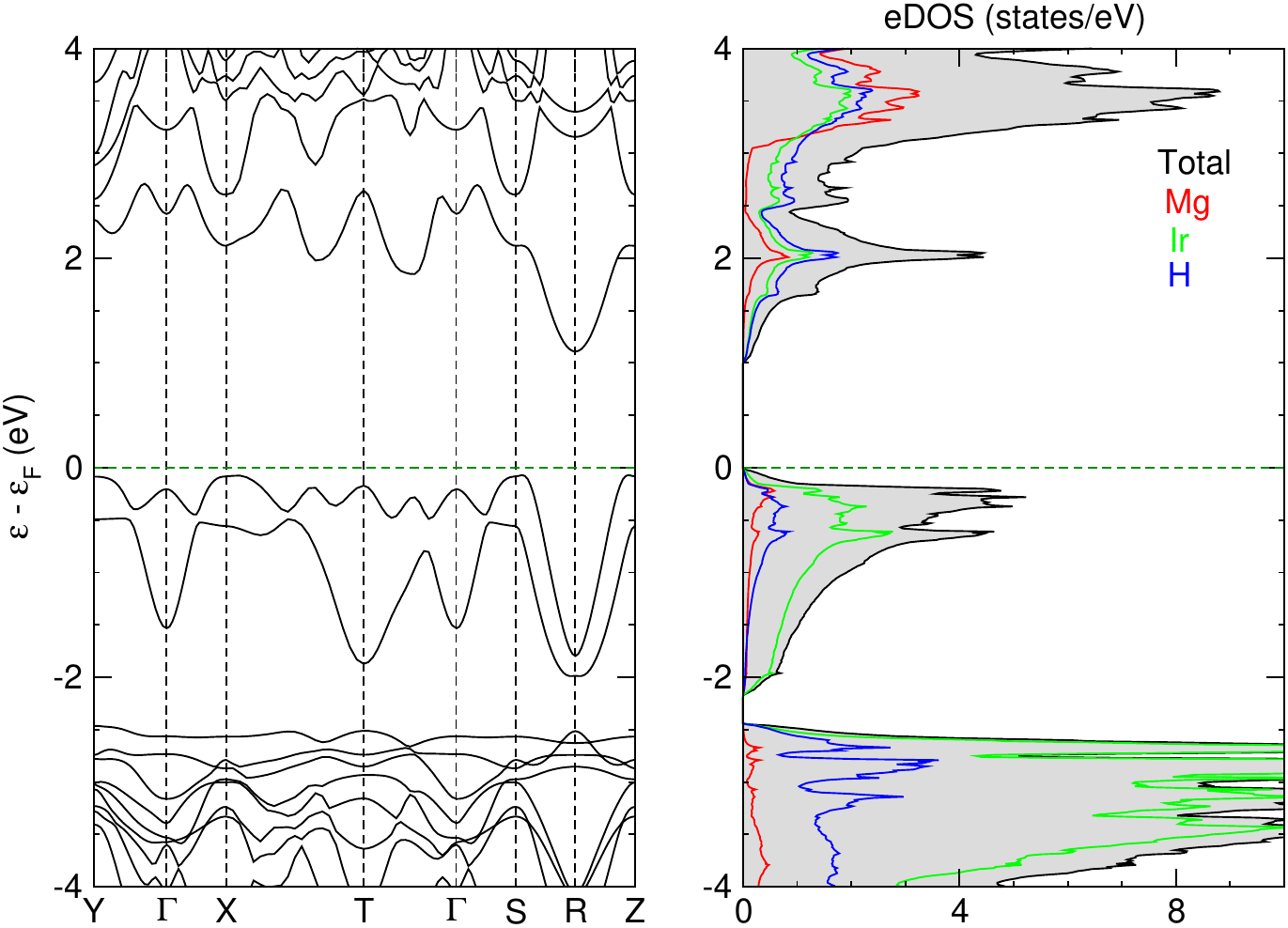}
\caption{In the left panel, electronic band structure of $Amm2$ Mg$_2$IrH$_6$ at
ambient pressure. In the right panel, the total electron density of states (eDOSs) projected into Mg (red), Ir (green), and H (blue) atoms.}
\label{fig:Amm2}
\end{figure}
\clearpage

\section{ Phonon mode-resolved $\lambda$}
\begin{figure}[h!]
\centering
\includegraphics[width=0.85\textwidth]{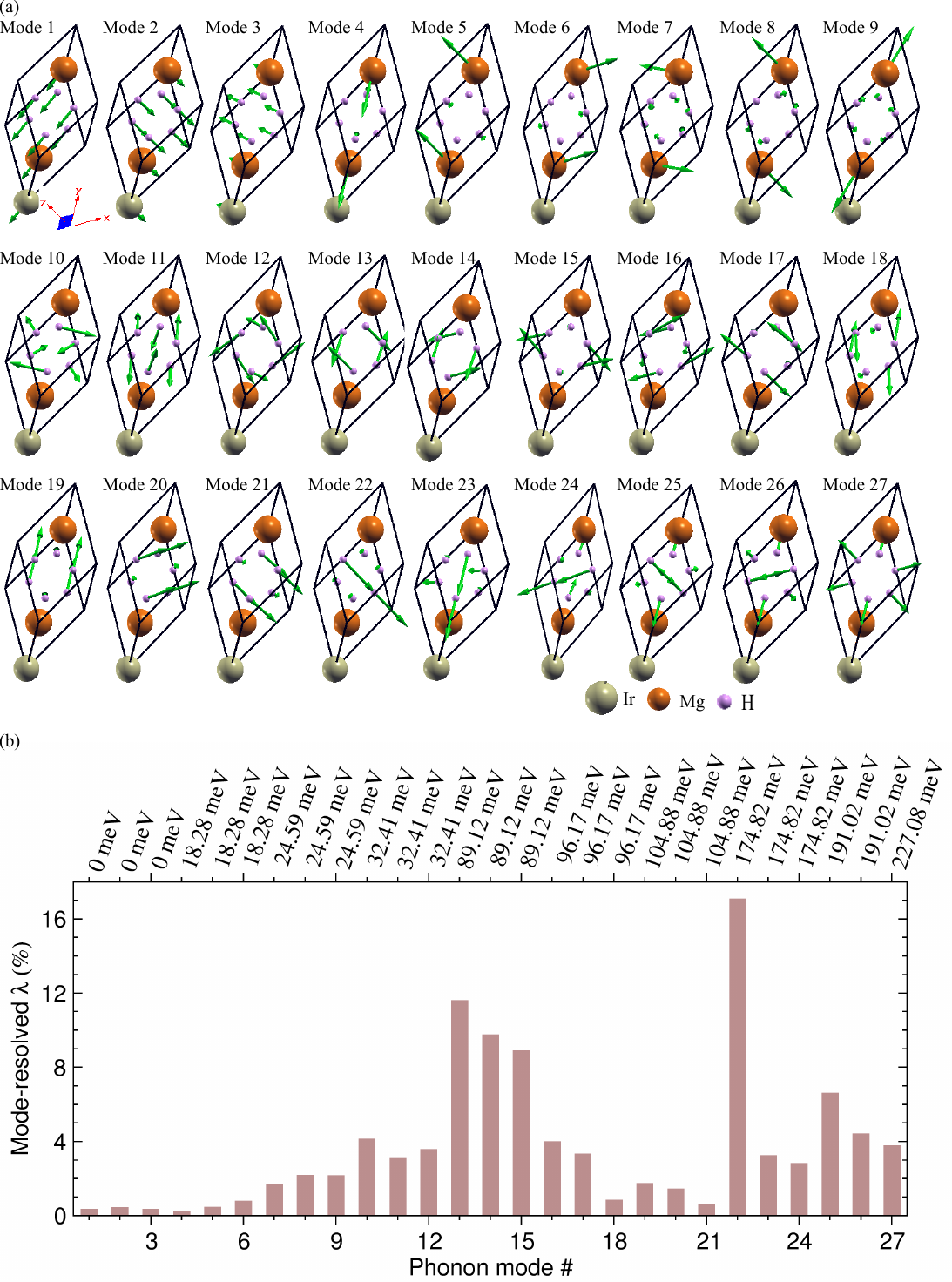}
\caption{ (a) Visualization of phonon eigenmodes at the $\Gamma$ point, and (b) mode-resolved $\lambda$. $x$-axis labels indicate the phonon mode index with their eigenvalues at the $\Gamma$ point.}
\label{fig:phmodes}
\end{figure}
\clearpage

\section{Electronic and phonon dispersion calculation using DFT+$U$}
\begin{figure}[h!]
\centering
\includegraphics[width=0.85\textwidth]{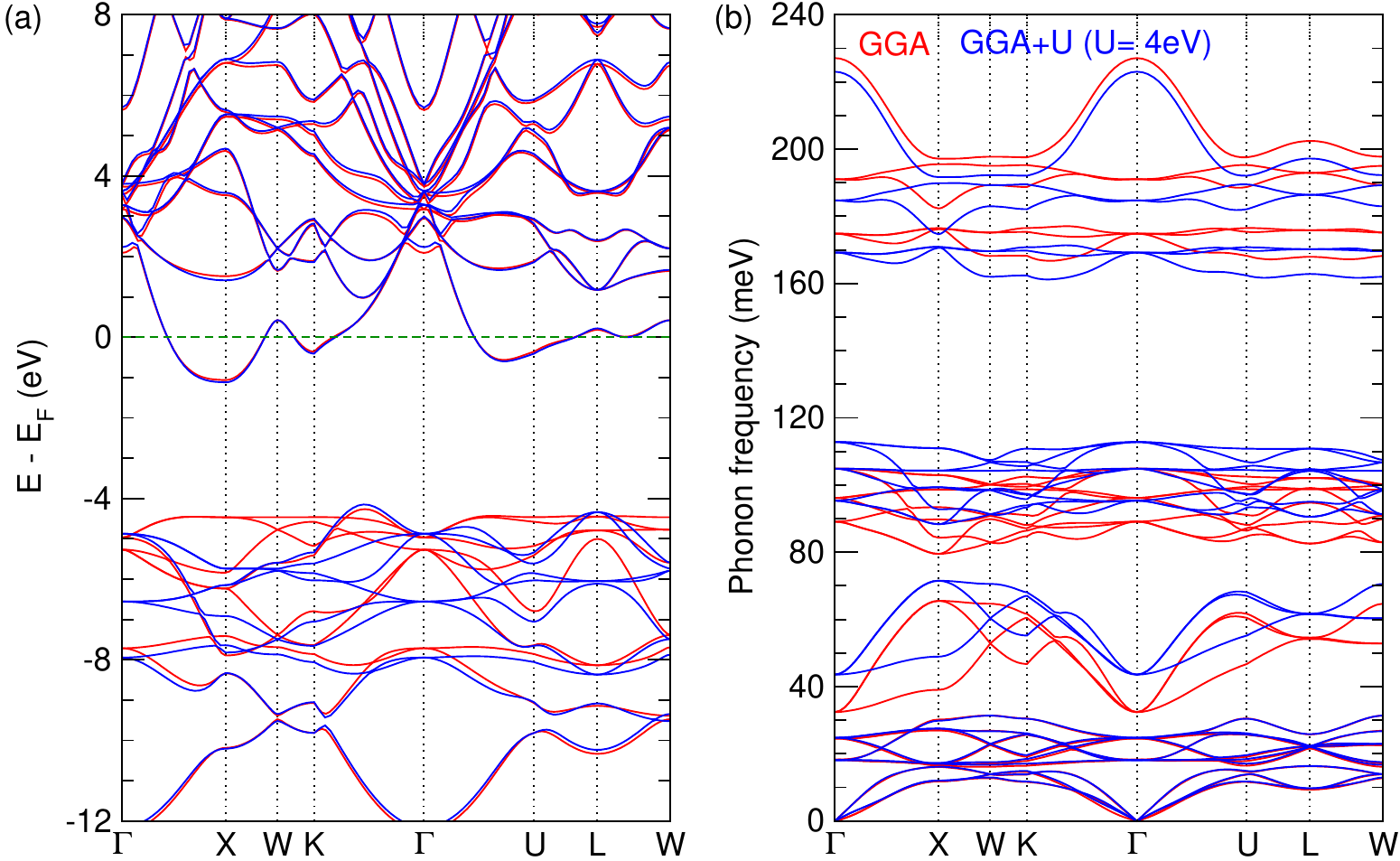}
\caption{ (a) Electron and (b) phonon dispersion calculated using GGA- (red) and GGA+$U$ (blue)-\textsc{XC} functionals. Hubbard $U$ ($U$= 4 eV) correction is included in correlated Ir-$d$ orbitals using the method of Cococcioni and de Gironcoli~\cite{Cococcioni2005} as implemented in \textsc{QE} package.}
\label{fig:e-ph_bands_ggau}
\end{figure}
\clearpage

\section{Free energy Calculations}
\begin{figure}[hb]
    \centering
    \includegraphics[width=0.5\textwidth]{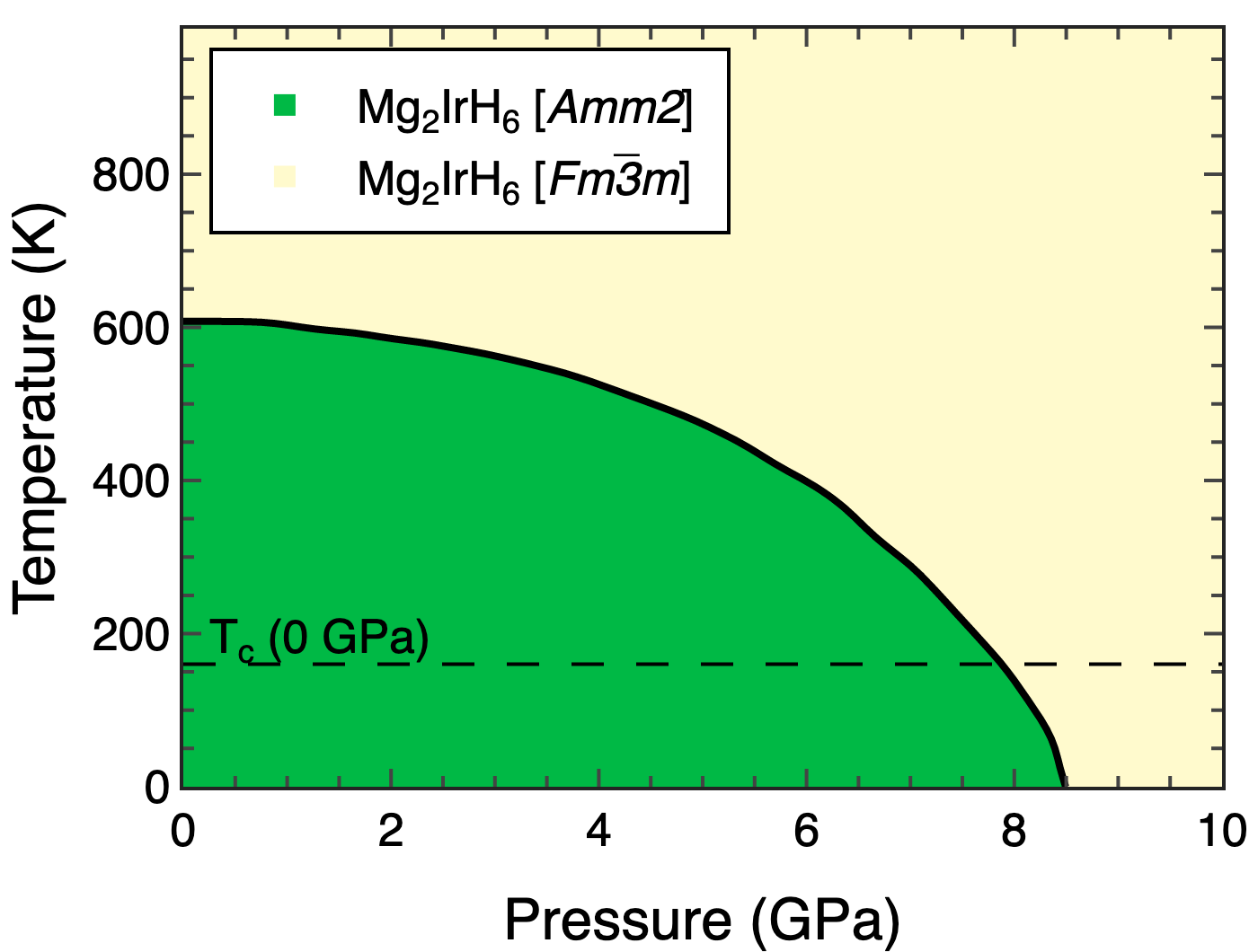}   
    \caption{Pressure-temperature phase diagram of $Amm2$ and $Fm\bar{3}m$ phases of Mg$_2$IrH$_6$ computed using quasi-harmonic free-energy calculations.}
    \label{fig:QHA}
\end{figure}
\clearpage
\section{Simulated Raman}
\begin{figure}[hb]
    \centering
    \includegraphics[width=0.9\linewidth]{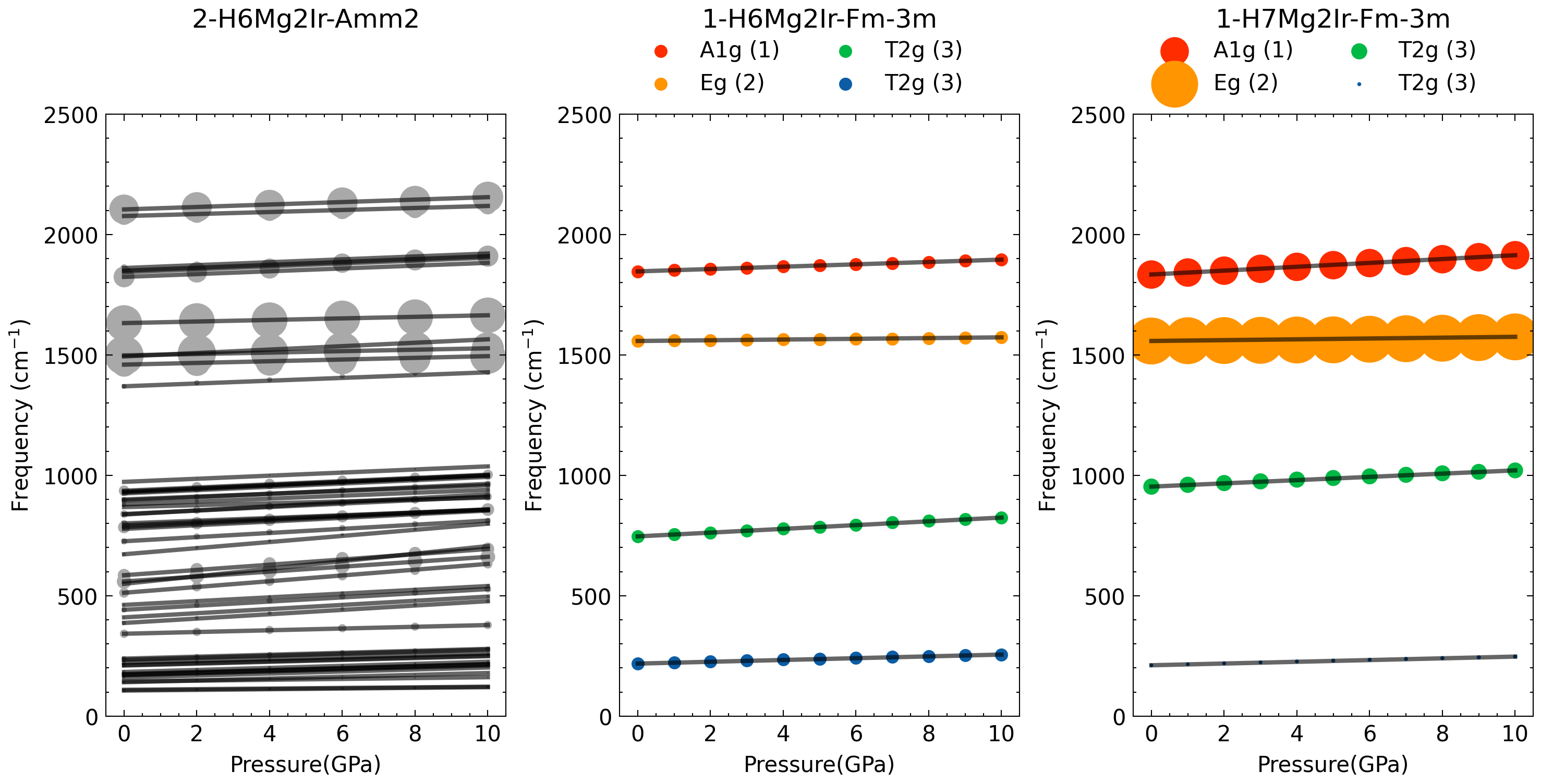}
    \caption{Calculated Raman\rev{-active} frequencies for Mg$_2$IrH$_6$ and Mg$_2$IrH$_7$ structures. Intensities for insulating compounds indicated by circle sizes. }
    \label{fig:raman}
\end{figure}
\begin{figure}[hb]
    \centering
    \includegraphics[width=0.9\linewidth]{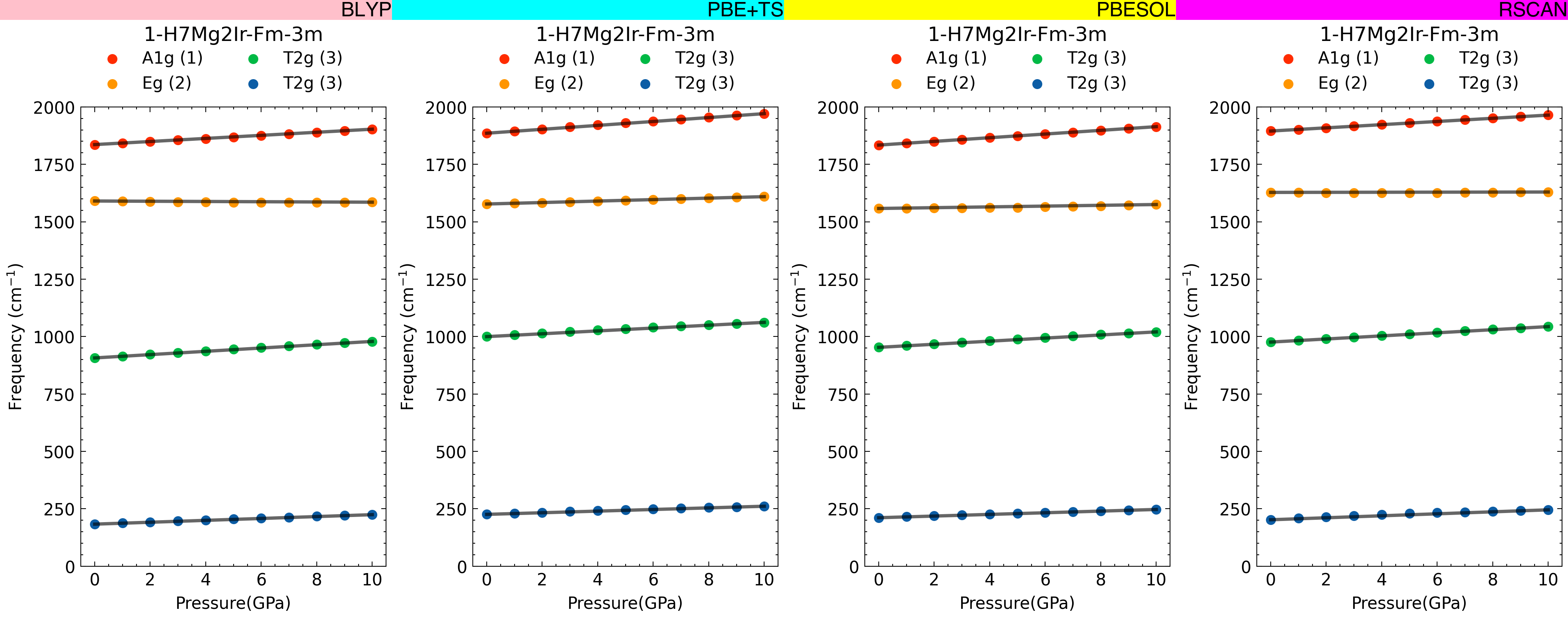}
    \caption{Simulated Raman\rev{-active} frequencies for cubic Mg$_2$IrH$_7$ for a range of exchange-correlation functionals. }
    \label{fig:raman-func}
\end{figure}
\clearpage

\section{\bf Anharmonic phonon calculations}
\begin{figure}[h!]
\centering
\includegraphics[width=0.75\textwidth]{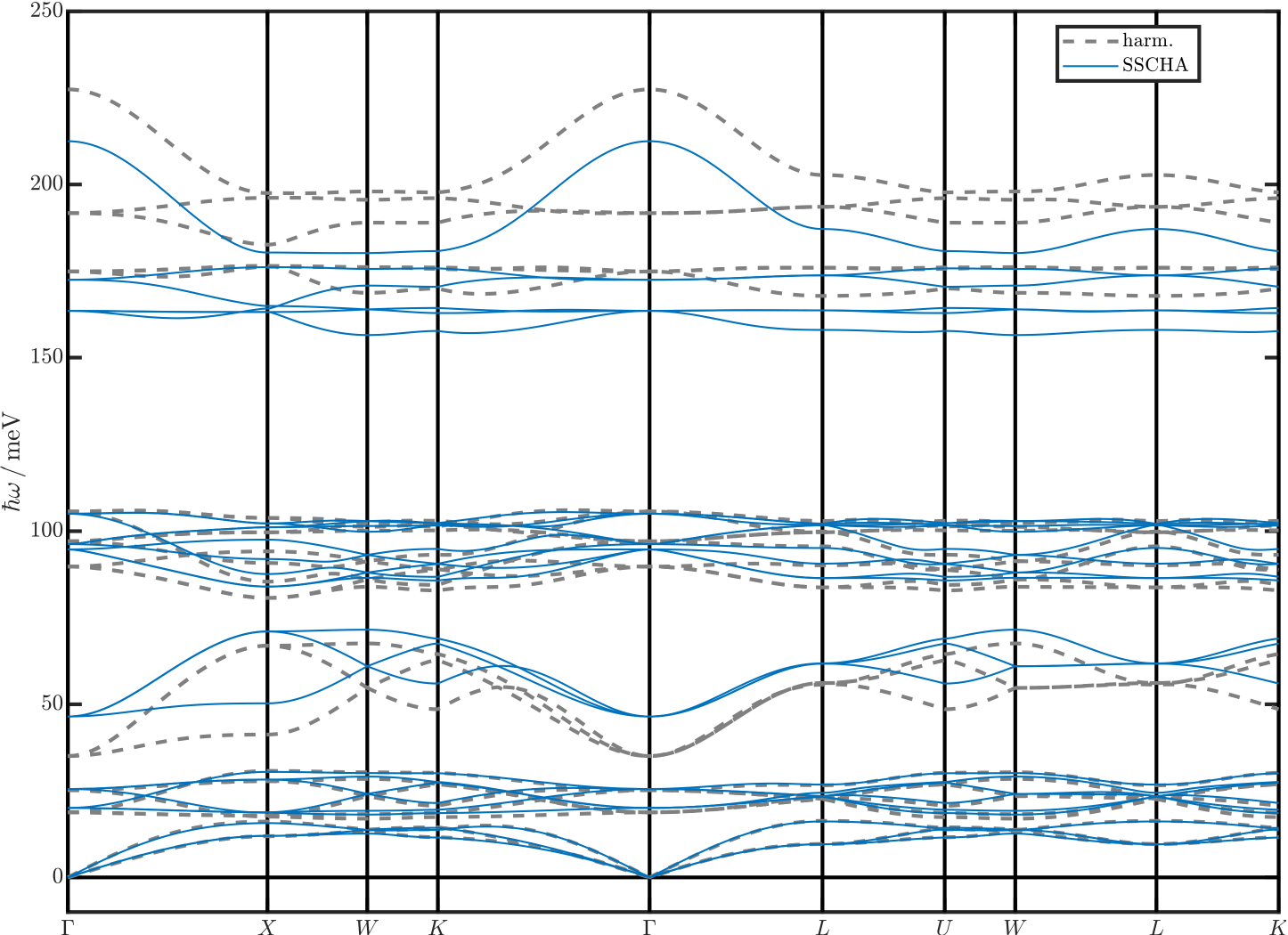}
\caption{Phonon dispersion of Mg$_2$IrH$_6$ including quantum anharmonic effects at 0\,K.}
\label{fig:SSCHA}
\end{figure}

\clearpage
\section{Crystal Structures [.res format]}

\begin{lstlisting}[float=hb]
TITL MgIrH-1-Mg2IrH6-Fm-3m 0.084500000000 71.835913608101 -15487.932800000000  0.00  0.00 9 (Fm-3m) n - 1
REM Functional PBE for solids (2008) Relativity Koelling-Harmon Dispersion off
REM Cut-off 600.0000 eV Grid scale 1.7500 Gmax 21.9610 1/A FBSC none
REM Offset 0.000 0.000 0.000 No. kpts 250 Spacing 0.03
REM MgIrH_0p0.cell (335d2f979913d52602ffaf9ace29feca)
REM Mg 3|1.8|7|8|9|20U:30:21:32
REM Ir 3|2.4|9|10|11|50U:60:51:52:43(qc=6)
REM H 1|0.6|13|15|17|10(qc=8)
CELL 1.54180    4.66608    4.66608    4.66608   60.00000   60.00000   60.00000
LATT -1
SFAC H  Mg Ir 
H      1  0.2398024272600  0.2398024272600  0.7601975727400 1.0
H      1  0.2398024272600  0.7601975727400  0.7601975727400 1.0
H      1  0.7601975727400  0.2398024272600  0.7601975727400 1.0
H      1  0.7601975727400  0.7601975727400  0.2398024272600 1.0
H      1  0.7601975727400  0.2398024272600  0.2398024272600 1.0
H      1  0.2398024272600  0.7601975727400  0.2398024272600 1.0
Mg     2  0.7500000000000  0.7500000000000  0.7500000000000 1.0
Mg     2  0.2500000000000  0.2500000000000  0.2500000000000 1.0
Ir     3  0.5000000000000  0.5000000000000  0.5000000000000 1.0
END
\end{lstlisting}

\begin{lstlisting}[float=hb]
TITL MgIrH-1-Mg2IrH7-Fm-3m 0.022400000000 72.344217934844 -15503.340800000000  0.00  0.00 10 (Fm-3m) n - 1
REM Functional PBE for solids (2008) Relativity Koelling-Harmon Dispersion off
REM Cut-off 600.0000 eV Grid scale 1.7500 Gmax 21.9610 1/A FBSC none
REM MP grid 9 9 9 Offset 0.000 0.000 0.000 No. kpts 35 Spacing 0.03
REM MgIrH_0p0.cell (335d2f979913d52602ffaf9ace29feca)
REM Mg 3|1.8|7|8|9|20U:30:21:32
REM Ir 3|2.4|9|10|11|50U:60:51:52:43(qc=6)
REM H 1|0.6|13|15|17|10(qc=8)
CELL 1.54180    4.67706    4.67706    4.67706   60.00000   60.00000   60.00000
LATT -1
SFAC H  Mg Ir 
H      1  0.2409774004889  0.7590225995111  0.2409774004889 1.0
H      1  0.2409774004889  0.2409774004889  0.7590225995111 1.0
H      1  0.7590225995111  0.2409774004889  0.2409774004889 1.0
H      1  0.7590225995111  0.2409774004889  0.7590225995111 1.0
H      1  0.7590225995111  0.7590225995111  0.2409774004889 1.0
H      1  0.2409774004889  0.7590225995111  0.7590225995111 1.0
H      1  0.0000000000000  0.0000000000000  0.0000000000000 1.0
Mg     2  0.2500000000000  0.2500000000000  0.2500000000000 1.0
Mg     2  0.7500000000000  0.7500000000000  0.7500000000000 1.0
Ir     3  0.5000000000000  0.5000000000000  0.5000000000000 1.0
END
\end{lstlisting}

\begin{lstlisting}[float=hb]
TITL MgIrH-1-Mg2IrH7-R-3m 0.025500000000 100.588817128868 -15504.146900000000  0.00  0.00 10 (R-3m) n - 1
REM Functional PBE for solids (2008) Relativity Koelling-Harmon Dispersion off
REM Cut-off 600.0000 eV Grid scale 1.7500 Gmax 21.9610 1/A FBSC none
REM MP grid 9 9 7 Offset 0.000 0.000 0.000 No. kpts 160 Spacing 0.03
REM MgIrH_0p0.cell (335d2f979913d52602ffaf9ace29feca)
REM Mg 3|1.8|7|8|9|20U:30:21:32
REM Ir 3|2.4|9|10|11|50U:60:51:52:43(qc=6)
REM H 1|0.6|13|15|17|10(qc=8)
CELL 1.54180    6.14379    6.14379    6.14379   43.72909   43.72909   43.72909
LATT -1
SFAC H  Mg Ir 
H      1  0.1026868854988  0.1026868854988  0.6090901528379 1.0
H      1  0.5000000000000  0.5000000000000  0.5000000000000 1.0
H      1  0.8973131145012  0.8973131145012  0.3909098471621 1.0
H      1  0.1026868854988  0.6090901528379  0.1026868854988 1.0
H      1  0.8973131145012  0.3909098471621  0.8973131145012 1.0
H      1  0.6090901528379  0.1026868854988  0.1026868854988 1.0
H      1  0.3909098471621  0.8973131145012  0.8973131145012 1.0
Mg     2  0.3923887698750  0.3923887698750  0.3923887698750 1.0
Mg     2  0.6076112301250  0.6076112301250  0.6076112301250 1.0
Ir     3 -0.0000000000000 -0.0000000000000  0.0000000000000 1.0
END
\end{lstlisting}

\begin{lstlisting}[float=hb]
TITL MgIrH-1-Mg3Ir-Fm-3m -0.003400000000 67.568564974623 -17079.898949999999  0.00  0.00 4 (Fm-3m) n - 1
REM Functional PBE for solids (2008) Relativity Koelling-Harmon Dispersion off
REM Cut-off 600.0000 eV Grid scale 1.7500 Gmax 21.9610 1/A FBSC none
REM MP grid 9 10 5 Offset 0.000 0.000 0.000 No. kpts 225 Spacing 0.03
REM MgIrH_0p0.cell (335d2f979913d52602ffaf9ace29feca)
REM Mg 3|1.8|7|8|9|20U:30:21:32
REM Ir 3|2.4|9|10|11|50U:60:51:52:43(qc=6)
CELL 1.54180    4.57179    4.57179    4.57179   60.00000   60.00000   60.00000
LATT -1
SFAC Mg Ir 
Mg     1  0.7500000000000  0.7500000000000  0.7500000000000 1.0
Mg     1  0.2500000000000  0.2500000000000  0.2500000000000 1.0
Mg     1  0.5000000000000  0.5000000000000  0.5000000000000 1.0
Ir     2  0.0000000000000  0.0000000000000  0.0000000000000 1.0
END
\end{lstlisting}

\begin{lstlisting}[float=hb]
TITL MgIrH-1-Mg3Ir2H7-C2m 0.067700000000 106.193022460744 -29209.100200000001  0.00  0.00 12 (C2/m) n - 1
REM Functional PBE for solids (2008) Relativity Koelling-Harmon Dispersion off
REM Cut-off 600.0000 eV Grid scale 1.7500 Gmax 21.9610 1/A FBSC none
REM Offset 0.000 0.000 0.000 No. kpts 297 Spacing 0.03
REM MgIrH_0p0.cell (335d2f979913d52602ffaf9ace29feca)
REM Mg 3|1.8|7|8|9|20U:30:21:32
REM Ir 3|2.4|9|10|11|50U:60:51:52:43(qc=6)
REM H 1|0.6|13|15|17|10(qc=8)
CELL 1.54180    4.69978    4.69978    5.61738   91.74118   91.74118   58.91363
LATT -1
SFAC H  Mg Ir 
H      1  0.3468593555633  0.8415574556774  0.6641452764272 1.0
H      1  0.3535461939612  0.3535461939612  0.6782277793160 1.0
H      1  0.0000000000000  0.0000000000000  0.0000000000000 1.0
H      1  0.8415574556774  0.3468593555633  0.6641452764272 1.0
H      1  0.6464538060388  0.6464538060388  0.3217722206840 1.0
H      1  0.1584425443226  0.6531406444367  0.3358547235728 1.0
H      1  0.6531406444367  0.1584425443226  0.3358547235728 1.0
Mg     2  0.1683588134073  0.1683588134073  0.3292901708938 1.0
Mg     2  0.8316411865927  0.8316411865927  0.6707098291062 1.0
Mg     2  0.5000000000000  0.5000000000000  0.0000000000000 1.0
Ir     3  0.1822869262993  0.1822869262993  0.8319899301768 1.0
Ir     3  0.8177130737007  0.8177130737007  0.1680100698232 1.0
END
\end{lstlisting}

\begin{lstlisting}[float=hb]
TITL MgIrH-1-Mg4Ir3H6-Im-3m -0.048200000000 129.091873959026 -42897.715199999999  0.00  0.00 13 (Im-3m) n - 1
REM Functional PBE for solids (2008) Relativity Koelling-Harmon Dispersion off
REM Cut-off 600.0000 eV Grid scale 1.7500 Gmax 21.9610 1/A FBSC none
REM MP grid 8 8 8 Offset 0.000 0.000 0.000 No. kpts 26 Spacing 0.03
REM MgIrH_0p0.cell (335d2f979913d52602ffaf9ace29feca)
REM Mg 3|1.8|7|8|9|20U:30:21:32
REM Ir 3|2.4|9|10|11|50U:60:51:52:43(qc=6)
REM H 1|0.6|13|15|17|10(qc=8)
CELL 1.54180    5.51451    5.51451    5.51451  109.47122  109.47122  109.47122
LATT -1
SFAC H  Mg Ir 
H      1  0.7600830742351  0.7600830742351  0.0000000000000 1.0
H      1  0.7600830742351  0.0000000000000  0.7600830742351 1.0
H      1  0.2399169257649  0.2399169257649  0.0000000000000 1.0
H      1  0.0000000000000  0.2399169257649  0.2399169257649 1.0
H      1  0.0000000000000  0.7600830742351  0.7600830742351 1.0
H      1  0.2399169257649  0.0000000000000  0.2399169257649 1.0
Mg     2  0.5000000000000  0.5000000000000  0.5000000000000 1.0
Mg     2  0.0000000000000  0.5000000000000  0.0000000000000 1.0
Mg     2  0.0000000000000  0.0000000000000  0.5000000000000 1.0
Mg     2  0.5000000000000  0.0000000000000  0.0000000000000 1.0
Ir     3  0.5000000000000  0.5000000000000  0.0000000000000 1.0
Ir     3  0.5000000000000  0.0000000000000  0.5000000000000 1.0
Ir     3  0.0000000000000  0.5000000000000  0.5000000000000 1.0
END
\end{lstlisting}

\begin{lstlisting}[float=hb]
TITL MgIrH-2-Mg2IrH5-Cmcm -0.076100000000 143.333691455769 -30945.621200000001  0.00  0.00 16 (Cc) n - 1
REM Functional PBE for solids (2008) Relativity Koelling-Harmon Dispersion off
REM Cut-off 600.0000 eV Grid scale 1.7500 Gmax 21.9610 1/A FBSC none
REM MP grid 8 8 6 Offset 0.000 0.000 0.000 No. kpts 108 Spacing 0.03
REM MgIrH_0p0.cell (335d2f979913d52602ffaf9ace29feca)
REM Mg 3|1.8|7|8|9|20U:30:21:32
REM Ir 3|2.4|9|10|11|50U:60:51:52:43(qc=6)
REM H 1|0.6|13|15|17|10(qc=8)
CELL 1.54180    4.74533    4.74533    6.38388   90.00666   90.00666   94.37834
LATT -1
SFAC H  Mg Ir 
H      1  0.2468522548024  0.2464521516470  0.7325950185302 1.0
H      1  0.2464521516470  0.2468522548024  0.2325950185302 1.0
H      1  0.4913546118048  0.9897566895505  0.9729166541177 1.0
H      1  0.9897566895505  0.4913546118048  0.4729166541177 1.0
H      1  0.4929055126189  0.9915437222381  0.4927685234535 1.0
H      1  0.9915437222381  0.4929055126189  0.9927685234535 1.0
H      1  0.2374952057684  0.7588304057632  0.7320830804846 1.0
H      1  0.7588304057632  0.2374952057684  0.2320830804846 1.0
H      1  0.7339803669797  0.2347279796414  0.7336850529610 1.0
H      1  0.2347279796414  0.7339803669797  0.2336850529610 1.0
Mg     2  0.9780943536533  0.9780914936485  0.4829233849779 1.0
Mg     2  0.9780914936485  0.9780943536533  0.9829233849779 1.0
Mg     2  0.5060649297031  0.5059498875726  0.9830398256762 1.0
Mg     2  0.5059498875726  0.5060649297031  0.4830398256762 1.0
Ir     3  0.4824950418691  0.0007279927391  0.2328381597988 1.0
Ir     3  0.0007279927391  0.4824950418691  0.7328381597988 1.0
END
\end{lstlisting}

\begin{lstlisting}[float=hb]
TITL MgIrH-2-Mg2IrH6-Amm2 0.007200000000 144.379851302026 -30976.059300000001  0.00  0.00 18 (Amm2) n - 1
REM Functional PBE for solids (2008) Relativity Koelling-Harmon Dispersion off
REM Cut-off 600.0000 eV Grid scale 1.7500 Gmax 21.9610 1/A FBSC none
REM MP grid 6 6 6 Offset 0.000 0.000 0.000 No. kpts 54 Spacing 0.03
REM MgIrH.cell (335d2f979913d52602ffaf9ace29feca)
REM Mg 3|1.8|7|8|9|20U:30:21:32
REM Ir 3|2.4|9|10|11|50U:60:51:52:43(qc=6)
REM H 1|0.6|13|15|17|10(qc=8)
CELL 1.54180    6.56232    4.69156    4.69156   88.32611   90.00000   90.00000
LATT -1
SFAC H  Mg Ir 
H      1  0.5000000000000  0.7560855173187  0.2485385147120 1.0
H      1  0.7597352680519  0.4997875636871  0.4997875636871 1.0
H      1  0.5000000000000  0.2408220253033  0.2408220253033 1.0
H      1  0.5000000000000  0.2485385147120  0.7560855173187 1.0
H      1  0.7464616102291  0.0036899515095  0.0036899515095 1.0
H      1  0.0000000000000  0.2554796300315  0.2554796300315 1.0
H      1  0.0000000000000  0.2647507829720  0.7376982665770 1.0
H      1  0.2535383897709  0.0036899515095  0.0036899515095 1.0
H      1  0.0000000000000  0.7473689432757  0.7473689432757 1.0
H      1  0.0000000000000  0.7376982665770  0.2647507829720 1.0
H      1  0.2402647319481  0.4997875636871  0.4997875636871 1.0
H      1  0.0000000000000  0.5068998641674  0.5068998641674 1.0
Mg     2  0.7429833434990  0.9881309589843  0.5050406993662 1.0
Mg     2  0.7429833434990  0.5050406993662  0.9881309589843 1.0
Mg     2  0.2570166565010  0.5050406993662  0.9881309589843 1.0
Mg     2  0.2570166565010  0.9881309589843  0.5050406993662 1.0
Ir     3  0.5000000000000  0.4891841959690  0.4891841959690 1.0
Ir     3  0.0000000000000  0.0021392434802  0.0021392434802 1.0
END
\end{lstlisting}

\begin{lstlisting}[float=hb]
TITL MgIrH-2-Mg2IrH9-P42m -0.009200000000 183.563587013802 -31069.982700000000  0.00  0.00 24 (P42/m) n - 1
REM Functional PBE for solids (2008) Relativity Koelling-Harmon Dispersion off
REM Cut-off 600.0000 eV Grid scale 1.7500 Gmax 21.9610 1/A FBSC none
REM MP grid 8 6 6 Offset 0.000 0.000 0.000 No. kpts 36 Spacing 0.03
REM MgIrH_0p0.cell (335d2f979913d52602ffaf9ace29feca)
REM Mg 3|1.8|7|8|9|20U:30:21:32
REM Ir 3|2.4|9|10|11|50U:60:51:52:43(qc=6)
REM H 1|0.6|13|15|17|10(qc=8)
CELL 1.54180    6.39037    6.39037    4.49505   90.00000   90.00000   90.00000
LATT -1
SFAC H  Mg Ir 
H      1  0.0580813378979  0.9842036057895  0.5000000000000 1.0
H      1  0.9419186621021  0.0157963942105  0.5000000000000 1.0
H      1  0.9842036057895  0.9419186621021  0.0000000000000 1.0
H      1  0.0157963942105  0.0580813378979  0.0000000000000 1.0
H      1  0.6742339054048  0.1921793668453  0.5000000000000 1.0
H      1  0.3257660945952  0.8078206331547  0.5000000000000 1.0
H      1  0.1921793668453  0.3257660945952  0.0000000000000 1.0
H      1  0.8078206331547  0.6742339054048  0.0000000000000 1.0
H      1  0.3470353706737  0.1034854487123  0.7604983732390 1.0
H      1  0.6529646293263  0.8965145512877  0.7604983732390 1.0
H      1  0.1034854487123  0.6529646293263  0.2604983732390 1.0
H      1  0.8965145512877  0.3470353706737  0.2604983732390 1.0
H      1  0.3470353706737  0.1034854487123  0.2395016267610 1.0
H      1  0.6529646293263  0.8965145512877  0.2395016267610 1.0
H      1  0.1034854487123  0.6529646293263  0.7395016267610 1.0
H      1  0.8965145512877  0.3470353706737  0.7395016267610 1.0
H      1  0.5000000000000  0.5000000000000  0.2500000000000 1.0
H      1  0.5000000000000  0.5000000000000  0.7500000000000 1.0
Mg     2  0.2611733307333  0.3824267486473  0.5000000000000 1.0
Mg     2  0.7388266692667  0.6175732513527  0.5000000000000 1.0
Mg     2  0.3824267486473  0.7388266692667  0.0000000000000 1.0
Mg     2  0.6175732513527  0.2611733307333  0.0000000000000 1.0
Ir     3  0.5000000000000  0.0000000000000  0.5000000000000 1.0
Ir     3  0.0000000000000  0.5000000000000  0.0000000000000 1.0
END
\end{lstlisting}

\clearpage

\section{Full Convex Hull}
\begin{figure}[hb]
    \centering
    \includegraphics{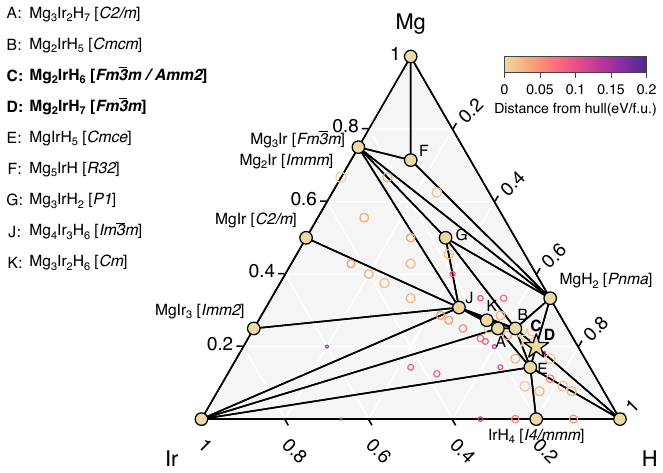} 
    \caption{Full Mg--Ir--H ternary convex hull at 20\,GPa.}
    \label{fig:convex}
\end{figure}
\clearpage

\section{Elastic Constants}
The elasticity of a lattice is described by its matrix of second-order elastic constants
\begin{equation}
    C_{ij} = \frac{1}{V_{o}} \frac{\partial^2 E}{\partial \epsilon_i\partial\epsilon_i},
\end{equation}
where $E$ is the energy of the crystal, $V_0$ its equilibrium volume, and $\epsilon$ denotes a strain. $C_{ij}$ also referred to as the stiffness constant, is a symmetric matrix of dimensions 6$\times$6. 

\begin{figure}[hb]
    \centering
    \includegraphics[width=0.5\linewidth]{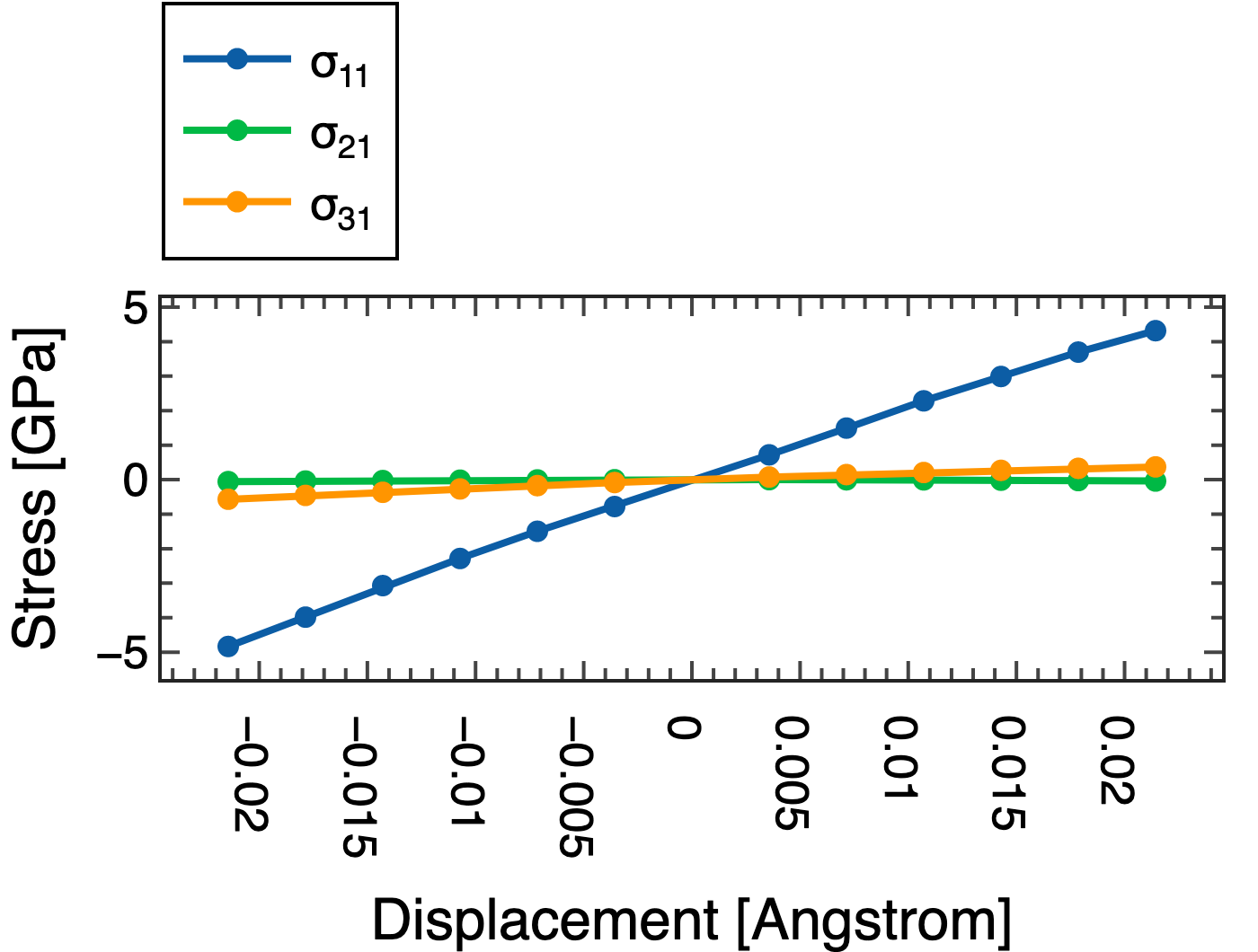}
    \caption{Stress-unit-cell displacement for cubic Mg$_2$Ir$_6$ at 0GPa, calculated using \textsc{castep}.}
    \label{fig:elastic}
\end{figure}

The ``Born elastic stability criteria"~\cite{Born1940} for mechanical stability is satisfied if, for a cubic structure,
\begin{equation}
C_{11} - C_{12} > 0,\quad C_{11} + 2C_{12} > 0,\quad\&\quad C_{44} > 0.
\end{equation}
Table~\ref{fig:elastic} shows that cubic Mg$_2$IrH$_6$ is indeed mechanically stable. 

\begin{table}[h!]
    \centering
    \caption{Elastic constants for cubic Mg$_2$IrH$_6$, calculated using \textsc{castep}.}
    \begin{tabular}{ll} \hline \hline
c$_{11}$ & 213.4$\pm$2.2\,GPa \\
c$_{12}$ &  55.0$\pm$2.6\,GPa \\
c$_{44}$ &  56.7$\pm$1.2\,GPa \\ \hline\hline
    \end{tabular}
    \label{tab:Elastic}
\end{table}

\clearpage

\section{Coarse T$_c$ calculations}
\subsection{High Symmetry Search}

\begin{table}[h!]
\caption{$T_{\rm c}$ calculations for the best structures appearing in the high-symmetry AIRSS search. * Energies are computed based on the high-symmetry AIRSS search. This means they may be knocked off the hull by a more thorough search, and so the energies are lower bounds. $T_{\rm c}$s are given with the electronic Gaussian broadening values given in parenthesis. We ranked the structures by both the smallest and largest broadening values to account for the poor convergence.  $^\circ$  Ca sites occupied by XY$_{12}$ icosahedra. $\dagger$: 1b, 3d, 3c Wyckoff sites occupied.}
    \begin{ruledtabular}
    \begin{tabular}{lllllll}
Composition         & Space Group & Prototype & $T{}^{\rm AD}_c$ (0.005 Ry) & $T{}^{\rm AD}_c$ (0.05Ry) & $\Delta H_f$\,(eV)* & $\Delta H$\,(eV)* \\ \hline
H$_{6}$Mg$_{2}$Ir   &  Fm-3m      & K$_2$PtCl$_6$       & 134.103         &  28.917  &  -0.263  &  \textbf{0.000}  \\
H$_{6}$NaAu$_{2}$   &  Fm-3m      & K$_2$PtCl$_6$       & 89.168          &  14.517  &  0.030   &  0.080           \\
H$_{3}$KPd          &  Pm-3m      & Cubic Perovskite    & 65.027          &  4.047   &  -0.224  &  0.026           \\
HNSc                &  F-43m      & Half-Heusler        & 38.765          &  19.333  &  -0.790  &  \textbf{0.000}  \\
H$_{3}$MgBa         &  Pm-3m      & Cubic Perovskite    & 35.503          &  21.422  &  -0.270  &  0.100           \\
H$_{12}$Lu$_{2}$Ta  &  Fm-3       & Fluorite$^\circ$     & 29.421          &  22.537  &  -0.248  &  0.075           \\
HAl$_{3}$Ir         &  Pm-3m      & Cubic Perovskite    & 27.011          &  10.822  &  -0.146  &  \textbf{0.000}  \\
H$_{3}$KPt          &  Pm-3m      & Cubic Perovskite    & 25.369          &  14.539  &  -0.281  &  \textbf{0.000}  \\
H$_{3}$NZr$_{3}$    &  Pm-3m      & $\dagger$                  & 12.871          &  10.292  &  -0.595  &  \textbf{0.000}  \\
HHf$_{3}$Pb         &  Pm-3m      & Cubic Perovskite    & 9.329           &  2.117   &  -0.117  &  0.031           \\
HNb$_{3}$O$_{3}$    &  Pm-3m      & $\dagger$           & 8.782           &  5.338   &  -1.854  &  \textbf{0.000}  \\
HMg$_{3}$Ag         &  Pm-3m      & Cubic Perovskite    & 8.442           &  5.266   &  -0.050  &  \textbf{0.000}  \\
HMgPd$_{3}$         &  Pm-3m      & Cubic Perovskite    & 6.821           &  0.007   &  -0.452  &  \textbf{0.000}  \\
HTaOs$_{3}$         &  Pm-3m      & Cubic Perovskite    & 6.415           &  3.682   &  -0.044  &  \textbf{0.000}  \\
HW$_{3}$Ir          &  Pm-3m      & Cubic Perovskite    & 4.350           &  4.563   &  0.085   &  0.085           \\
HNa$_{3}$Au         &  Pm-3m      & Cubic Perovskite    & 4.309           &  0.781   &  -0.146  &  \textbf{0.000}  \\
HAlRh$_{3}$         &  Pm-3m      & Cubic Perovskite    & 3.743           &  1.803   &  -0.350  &  \textbf{0.000}  \\
HAlHf$_{3}$         &  Pm-3m      & Cubic Perovskite    & 2.146           &  1.465   &  -0.287  &  \textbf{0.000}  \\
HN$_{3}$Mo$_{3}$    &  Pm-3m      & $\dagger$           & 2.118           &  4.153   &  -0.491  &  \textbf{0.000}  \\
HHf$_{3}$Tl         &  Pm-3m      & Cubic Perovskite    & 2.015           &  1.415   &  -0.122  &  0.026           \\
HHf$_{3}$Hg         &  Pm-3m      & Cubic Perovskite    & 1.222           &  1.000   &  -0.199  &  \textbf{0.000}  \\
H$_{3}$NaPd         &  Pm-3m      & Cubic Perovskite    & 0.639           &  0.038   &  -0.277  &  \textbf{0.000}  \\
HSiTa$_{3}$         &  Pm-3m      & Cubic Perovskite    & 0.000           &  3.114   &  -0.147  &  \textbf{0.000}  \\
\end{tabular}
\end{ruledtabular}
\end{table}

\clearpage
\begin{lstlisting}[float=hb]
TITL th-1-HNSc-F-43m 0.983600000000 30.077721864843 -1567.067620000000  0.00  0.00 3 (F-43m) n - 1
CELL 1.54180    3.49076    3.49076    3.49076   60.00000   60.00000   60.00000
LATT -1
SFAC H  N  Sc 
H      1  0.2500000000000  0.2500000000000  0.2500000000000 1.0
N      2  0.7500000000000  0.7500000000000  0.7500000000000 1.0
Sc     3  0.5000000000000  0.5000000000000  0.5000000000000 1.0
END
\end{lstlisting}

\begin{lstlisting}[float=hb]
TITL th-1-HN3Mo3-Pm-3m 1.010800000000 69.750856277710 -6443.849310000000  0.00  0.00 7 (Pm-3m) n - 1
CELL 1.54180    4.11639    4.11639    4.11639   90.00000   90.00000   90.00000
LATT -1
SFAC H  N  Mo 
H      1  0.5000000000000  0.5000000000000  0.5000000000000 1.0
N      2  0.5000000000000  0.0000000000000  0.0000000000000 1.0
N      2  0.0000000000000  0.0000000000000  0.5000000000000 1.0
N      2  0.0000000000000  0.5000000000000  0.0000000000000 1.0
Mo     3  0.5000000000000  0.0000000000000  0.5000000000000 1.0
Mo     3  0.5000000000000  0.5000000000000  0.0000000000000 1.0
Mo     3  0.0000000000000  0.5000000000000  0.5000000000000 1.0
END
\end{lstlisting}

\begin{lstlisting}[float=hb]
TITL th-1-H12Lu2Ta-Fm-3 1.051200000000 114.705611487894 -22614.203699999998  0.00  0.00 15 (Fm-3) n - 1
CELL 1.54180    5.45381    5.45381    5.45381   60.00000   60.00000   60.00000
LATT -1
SFAC H  Lu Ta 
H      1  0.5900714906738  0.1548830550478  0.8451169449522 1.0
H      1  0.1548830550478  0.8451169449522  0.5900714906738 1.0
H      1  0.5900714906738  0.4099285093262  0.1548830550478 1.0
H      1  0.4099285093262  0.8451169449522  0.1548830550478 1.0
H      1  0.8451169449522  0.1548830550478  0.4099285093262 1.0
H      1  0.4099285093262  0.5900714906738  0.8451169449522 1.0
H      1  0.8451169449522  0.5900714906738  0.1548830550478 1.0
H      1  0.1548830550478  0.4099285093262  0.8451169449522 1.0
H      1  0.1548830550478  0.5900714906738  0.4099285093262 1.0
H      1  0.8451169449522  0.4099285093262  0.5900714906738 1.0
H      1  0.4099285093262  0.1548830550478  0.5900714906738 1.0
H      1  0.5900714906738  0.8451169449522  0.4099285093262 1.0
Lu     2  0.7500000000000  0.7500000000000  0.7500000000000 1.0
Lu     2  0.2500000000000  0.2500000000000  0.2500000000000 1.0
Ta     3  0.5000000000000  0.5000000000000  0.5000000000000 1.0
END
\end{lstlisting}

\clearpage
\subsection{A$_2$BH$_6$ High throughput Screening}

\begin{table}[h!]
\caption{High-throughput screening of A$_2$BH$_6$ structures. * Energies are computed based on the high-symmetry AIRSS search. This means they may be knocked off the hull by a more thorough search, and so the energies are lower bounds.}
    \begin{ruledtabular}
    \begin{tabular}{llllll}
A & B in A$_2$BH$_6$ & $T{}^{\rm AD}_c$ (degauss = 0.005 Ry) & T${}^{AD}_c$ (degauss = 0.05Ry) & $\Delta H_f$\,(eV)* & $\Delta H$\,(eV)* \\ \hline
  Mg  &  Ir &  134.386  &  29.353  &  -0.263  &  \textbf{0.000} \\
  Mg  &  Pt &  105.533  &  86.189  &  -0.084  &  0.088 \\
  Al  &  Re &  105.119  &  30.859  &  0.001   &  0.001 \\
  Mg  &  Rh &  102.255  &  31.029  &  -0.248  &  \textbf{0.000} \\
  Na  &  Ag &  99.590   &  37.814  &  0.163   &  0.264 \\
  Al  &  Mn &  89.395   &  18.447  &  -0.011  &  \textbf{0.000} \\
  Na  &  Au &  77.823   &  26.251  &  -0.012  &  0.089 \\
  In  &  Re &  59.837   &  18.892  &  0.165   &  0.165 \\
  Ca  &  Ag &  58.098   &  31.324  &  -0.193  &  0.197 \\
  In  &  Tc &  57.933   &  20.296  &  0.133   &  0.143 \\
  Sr  &  Ag &  48.735   &  18.795  &  -0.223  &  0.156 \\
  Ba  &  Ag &  41.980   &  16.142  &  -0.233  &  0.101 \\
  Na  &  B &   41.031   &  24.301  &  0.117   &  0.218 \\
  Na  &  Zn &  39.785   &  44.523  &  0.808   &  0.909 \\
  Ca  &  Cu &  36.164   &  6.794   &  -0.321  &  0.069 \\
  Na  &  Al &  35.128   &  23.398  &  0.052   &  0.155 \\
  Ca  &  Pd &  34.038   &  11.898  &  -0.427  &  0.081 \\
  Na  &  Ga &  31.550   &  25.716  &  0.126   &  0.227 \\
  Lu  &  Tc &  29.302   &  1.658   &  -0.528  &  \textbf{0.000} \\
  Sr  &  Cu &  26.670   &  4.605   &  -0.329  &  0.051 \\
  Sr  &  Pd &  26.498   &  7.773   &  -0.437  &  \textbf{0.000} \\
  Ba  &  Na &  25.671   &  12.309  &  -0.102  &  0.283 \\
  Y   &  Cu &  24.478   &  9.846   &  -0.364  &  0.182 \\
  Lu  &  Re &  23.703   &  3.079   &  -0.474  &  0.023 \\
  Ca  &  Ga &  23.306   &  17.366  &  -0.179  &  0.211 \\
  Nb  &  Ni &  22.974   &  20.898  &  -0.042  &  0.148 \\
  Lu  &  Ir &  21.524   &  14.576  &  -0.385  &  0.119 \\
  Rb  &  Ag &  21.292   &  10.443  &  0.098   &  0.213 \\
  Cs  &  Ag &  19.290   &  9.751   &  0.081   &  0.209 \\
  Sr  &  Ga &  15.155   &  9.076   &  -0.230  &  0.149 \\
  K   &  Au &  13.937   &  4.462   &  -0.095  &  0.067 \\
\end{tabular}
\end{ruledtabular}
\end{table}

\clearpage


\end{document}